%% file: ANA-STDM-2017-28-PAPER.tex
\pdfoutput=1
\newcommand*{\ATLASLATEXPATH}{}
\documentclass[PAPER, atlasdraft=false, texlive=2016, UKenglish, cernpreprint]{\ATLASLATEXPATH atlasdoc}
 
\usepackage{\ATLASLATEXPATH atlaspackage}
\usepackage{adjustbox}
\usepackage{\ATLASLATEXPATH atlasbiblatex}
 
\usepackage{\ATLASLATEXPATH atlascontribute}
 
\usepackage[process]{\ATLASLATEXPATH atlasphysics}
 
\graphicspath{{logos/}{figures/}}
 
\usepackage{ANA-STDM-2017-28-PAPER-defs}

\AtlasTitle{Measurement of distributions sensitive to the underlying event in inclusive Z-boson production in pp collisions at $\sqrt{s}$ = 13 TeV with the ATLAS detector }
 
\AtlasVersion{1.0}
\AtlasAbstract{
This paper presents measurements of charged-particle distributions sensitive to the properties of the underlying event in events containing a \ensuremath{Z}\xspace boson decaying into a muon pair.
The data were obtained using the ATLAS detector at the LHC in proton--proton collisions at a centre-of-mass energy of 13~\ensuremath{\text{Te\kern -0.1em V}}
with an integrated luminosity of 3.2~\mbox{fb$^{-1}$}.
Distributions of the charged-particle multiplicity and of the charged-particle transverse momentum are measured in regions of the azimuth defined relative to the \ensuremath{Z}\xspace boson direction.
The measured distributions are compared with the predictions of various Monte Carlo generators which implement different underlying-event models.
The Monte Carlo model predictions qualitatively describe the data well, but with some significant discrepancies.
}
 
\author{The ATLAS Collaboration}
 
\AtlasRefCode{STDM-2017-28}
 
\PreprintIdNumber{CERN-EP-2019-064}

\AtlasJournalRef{Eur. Phys. J. C 79 (2019) 666}
\AtlasDOI{10.1140/epjc/s10052-019-7162-0}

\AtlasCoverEgroupEditors{atlas-STDM-2017-28-editors@cern.ch}
 
\AtlasCoverEgroupEdBoard{atlas-STDM-2017-28-editorial-board@cern.ch}

\hypersetup{pdftitle={ATLAS document},pdfauthor={The ATLAS Collaboration}}
 
\begin{document}
 
\maketitle
 
\tableofcontents

\section{Introduction}
 
\newcommand{\AtlasCoordFootnote}{
ATLAS uses a right-handed coordinate system with its origin at the nominal interaction point (IP)
in the centre of the detector and the $z$-axis along the beam pipe.
The $x$-axis points from the IP to the centre of the LHC ring,
and the $y$-axis points upwards.
Cylindrical coordinates $(r,\phi)$ are used in the transverse plane,
$\phi$ being the azimuthal angle around the $z$-axis.
The pseudorapidity is defined in terms of the polar angle $\theta$ as $\eta = -\ln \tan(\theta/2)$.
Angular distance is measured in units of $\Delta R \equiv \sqrt{(\Delta\eta)^{2} + (\Delta\phi)^{2}}$.}
 
A typical proton--proton ($pp$) collision studied at the LHC consists of a short-distance hard-scattering process and accompanying activity collectively termed the underlying event (UE).
The hard scattering processes have a momentum transfer sufficiently large that the strong coupling constant is small and the cross-section may be calculated perturbatively in quantum chromodynamics (QCD).
The driving mechanisms for the production of the \UE\ are at a much lower momentum scale. These mechanisms include partons not participating in the hard-scattering process (beam remnants), radiation processes and
additional hard and semi-hard scatters in the same $pp$ collision, termed multiple parton interactions (MPI).
Phenomenological models are required to describe these processes using several free parameters determined from experiment. In addition to furthering the understanding of the proton's internal structure and the related soft-QCD processes, accurate modelling of the \UE\ is crucial for many data analyses at a hadron collider, either to precisely determine Standard Model quantities or to search for new particles and interactions.
 
The \UE\ is not distinguishable from the hard scatter on an event-by-event basis.
However, there are observables which are sensitive to the \UE\ properties, as first introduced by the CDF Collaboration in proton--antiproton ($p\bar{p}$) collisions at a centre-of-mass energy of 1.8~\TeV~\cite{Affolder:2001xt}.
An example of such an observable can be defined by topological considerations, based on the activity measurement in the direction transverse \footnote{\AtlasCoordFootnote{}} to a reference object.
 
The object in the event with the leading transverse momentum relates the \UE\ activity to the scale of the momentum transfer in the hard interaction.
In general, processes with leptonic final states like Drell--Yan events are experimentally clean and theoretically well understood, allowing reliable identification of the particles from the \UE.  The absence of QCD final-state radiation (FSR) permits a study of different kinematic regions with varying transverse momenta of the \Zboson\ boson due to harder or softer initial-state radiation (ISR). \\
 
Previous measurements of distributions sensitive to the properties of the \UE\ in Drell--Yan events were performed in $pp$ collisions at a centre-of-mass energy of 7~\TeV{} by the ATLAS \cite{STDM-2011-42} and CMS \cite{CMS-QCD-11-012} Collaborations and at a centre-of-mass energy of 13~\TeV{} by the CMS Collaboration \cite{CMS-FSQ-16-008}.
Both measurements at \rootseven{} verified that the dependence of the UE activity on the dimuon invariant mass is qualitatively well described by the \Powheg{}+\Pythiaeight{} and \HERWIGpp{} sets of tuned parameters but with some significant discrepancies. Reference \cite{STDM-2011-42} provides distributions which are sensitive to the choice of parameters used in the various \UE\ models.
 
This paper presents distributions of four observables sensitive to the \UE\ in events containing a \Zboson{} boson produced in
$pp$ collisions at a centre-of-mass energy of 13~\TeV{} in the ATLAS detector at the LHC, where the singly produced \Zboson{} boson decays into $\mu^{+}\mu^{-}$.
Observables measured as a function of the transverse momentum of the \Zboson{} boson, \Zpt{}, in various regions of phase space are compared with predictions from several Monte Carlo (MC) event generators.

\section{Underlying-event observables and measurement strategy}
 
\begin{figure*}
\begin{center}
\begin{minipage}{0.32\textwidth}
\resizebox{1.0\textwidth}{!}{\includegraphics{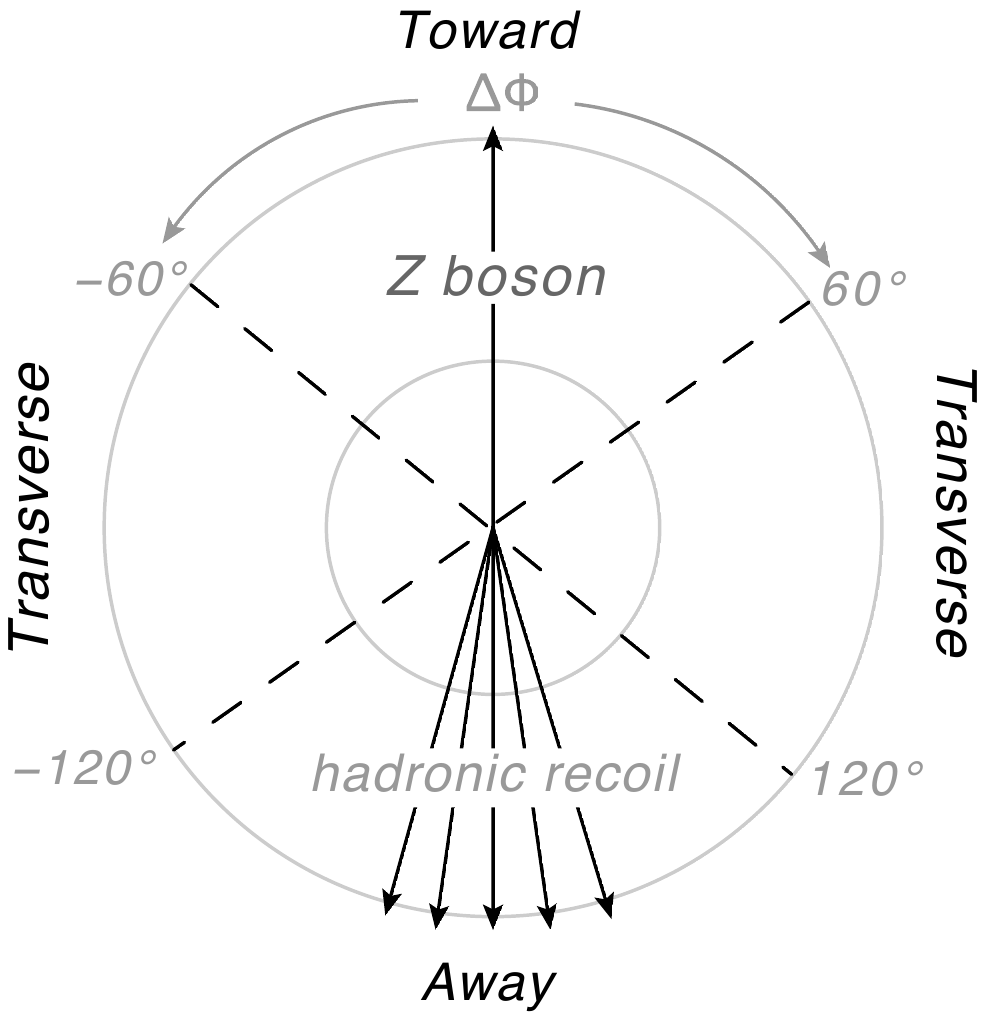}}
\end{minipage}
\hspace{0.1cm}
\begin{minipage}{0.65\textwidth}
\resizebox{1.0\textwidth}{!}{\includegraphics{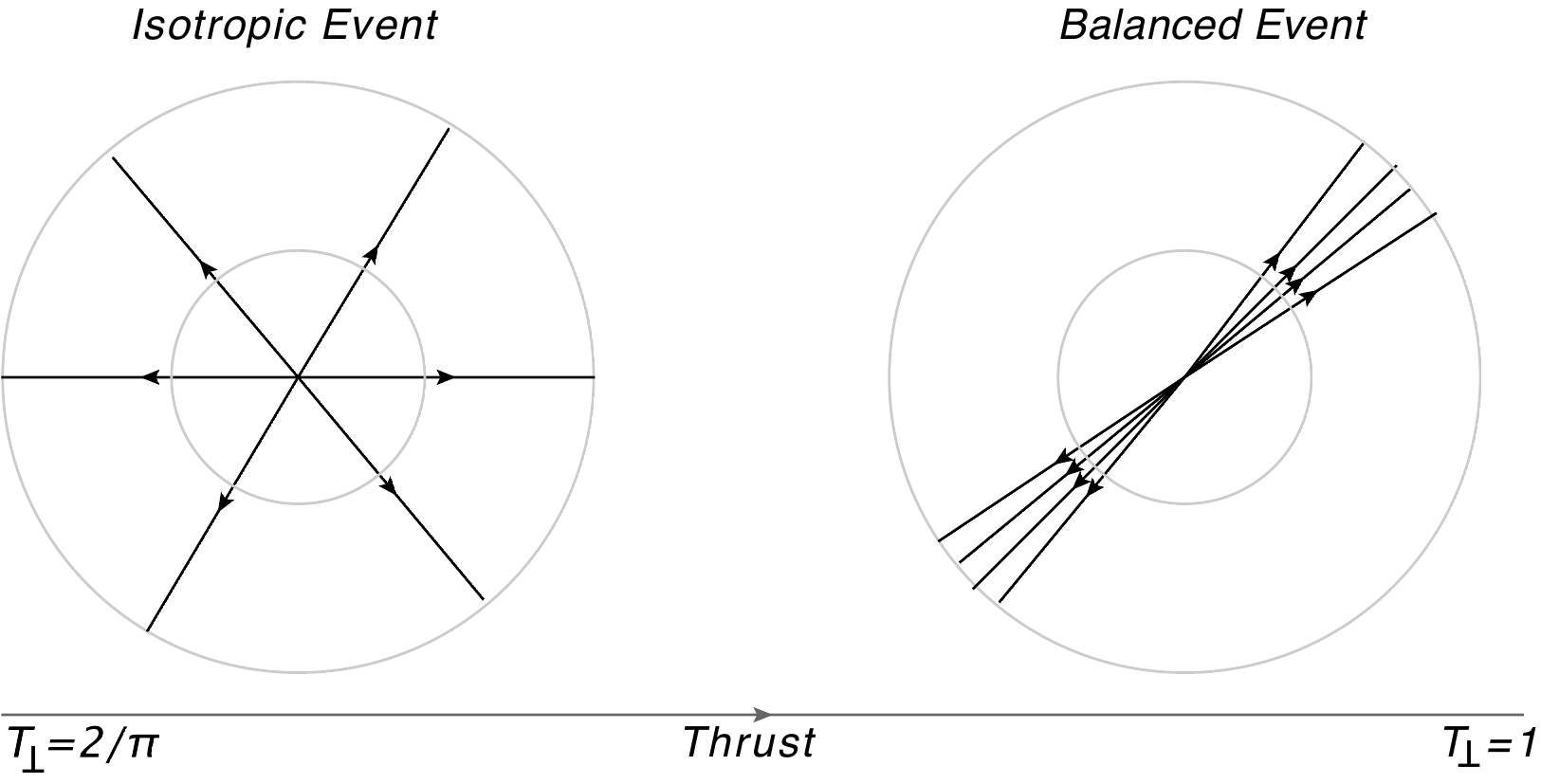}}
\end{minipage}
\end{center}
\caption{(a) Illustration of away, transverse, and towards regions in the transverse plane defined relative to the direction of the \Zboson{} boson.
(b) Illustration of an isotropic and a balanced event topology in the transverse plane with their corresponding values of thrust \thrust{}. In these figures, the beams are travelling perpendicular to the plane of the page.}
\label{fig:RegionsDefintion}
\end{figure*}

Events containing two muons originating from the decay of a singly produced \Zboson{} boson form a particularly interesting sample for studying the \UE. The final-state \Zboson{} boson is well-identified and colour neutral, so that interaction between the final-state leading particle and the \UE\ is minimal. Gluon radiation from the quarks or gluons initiating the hard scatter are, however, an important consideration as these give the remainder of the event a non-zero transverse momentum and change the kinematics of the final-state.
Observables are therefore measured in different regions of the transverse plane, which are defined relative to the direction of the \Zboson{} boson as illustrated in Figure~\ref{fig:RegionsDefintion}.
 
A charged particle lies in the \textit{away} region if its azimuthal angle relative to the \Zboson{} boson direction $|\Delta\phi|$ is greater than $120^\circ$. This region is heavily dominated by the hadronic recoil against the \Zboson{} boson from initial state quark/gluon radiation and is therefore not particularly sensitive to the \UE.
The \textit{toward} ($|\Delta\phi|\leq60^\circ$) and \textit{transverse} ($60^\circ<|\Delta\phi|\leq120^\circ$) regions contain less contamination from the hard process after subtraction of the two muons from the \Zboson{} boson.
The transverse region is sensitive to the \UE\ because, by construction, it is perpendicular to the direction of the \Zboson\ boson and hence is expected to have a lower level of activity from the hard-scattering process than the away region.
The two \textit{transverse} regions are differentiated on an event-by-event basis by their scalar sum of charged-particle \pT{}. The one with the larger sum is labelled \textit{trans-max} and the other \textit{trans-min}~\cite{Marchesini:1988hj,Pumplin:1997ix}.
The trans-min region is highly sensitive to the \UE\ activity because it is less likely that activity from recoiling jets leaks into this region.
 
Four distributions are studied to understand the \UE\ activity. The first is the charged-particle transverse momentum \ptspec{} distribution inclusive over all selected particles.
The final spectrum for this variable is accumulated over all events and then normalized.
The next three are evaluated on an event-by-event basis: the charged-particle multiplicity \dnch{}, the scalar sum of the transverse momentum of those particles \dsumpt{},
and the mean transverse momentum \dmeanpt{}, where mean \pT\ is the quotient of \sumpt{} and \nch{} (provided $\nch>0$ in the corresponding region).
The distributions of these variables are produced separately for charged particles lying in each of the regions described above.
The charged-particle multiplicity and the scalar sum of transverse momenta are normalized relative to the area of the corresponding region in the $\eta$--$\phi$ space.
This simplifies the comparison of the activity in different regions.
The distributions are distinguished in different ranges of the \Zboson{} boson transverse momentum \Zpt\ and for two regions of transverse thrust \thrust{} \cite{Banfi:2010xy}. Transverse thrust characterizes the topology of the tracks in the event and is
\begin{equation}
\thrust{} = \frac{ \sum_i | \vec{p_{\textrm{T},i}} \cdot \hat{n} | }{ \sum_i | \vec{p_{\textrm{T},i}} | }.
\end{equation}
The thrust axis $\hat{n}$ is the unit vector which maximizes \thrust{}. Here the summation is done on an event-by-event basis over the transverse momenta \pt{} of all charged particles except the two muons.
Transverse thrust has a maximum value of 1 for a pencil-like dijet topology and a minimum value of $2/\pi$ for a circularly symmetric distribution of particles in the transverse plane, as illustrated in Figure~\ref{fig:RegionsDefintion}. As proposed in Ref.~\cite{Kar:2018uno}, events with lower values of \thrust{} are more sensitive to the MPI component of the \UE. The two regions of thrust examined in this paper are $\thrust\leq0.75$ and $\thrust>0.75$, which are optimized to distinguish extra jet activity from the actual \UE\ activity. A measurement of transverse thrust in combination with the \UE\ activity was done at $\sqrt{s}=\SI{7}{\tera\electronvolt}$~\cite{STDM-2014-07}, but it did not distinguish the transverse regions.
 
In this paper, all measurements are also performed inclusively in $\thrust$. In total, the spectra of the four observables are measured in
96 regions of phase space, i.e. in eight bins of \Zpt{}; in the away, toward, trans-max, and trans-min regions; and for low, high, and inclusive \thrust{}.
The bin boundaries in \Zpt\ are $(0,10,20,40,60,80,120,200,500)$~\GeV.
In addition to distributions of the four observables, the arithmetic means $\langle\nch\rangle$, $\langle\sumpt\rangle$, and $\langle\meanpt\rangle$ are evaluated as functions of \Zpt\ in each of the various regions of phase space.
 
\section{The ATLAS detector}
 
The ATLAS detector~\cite{PERF-2007-01, ATLAS-TDR-2010-19, PIX-2018-001} at the LHC covers nearly the entire solid angle around the collision point.
It consists of an inner tracking detector (ID) surrounded by a thin superconducting solenoid, electromagnetic and hadronic calorimeters,
and a muon spectrometer (MS) incorporating three large superconducting toroid magnets.
 
The ID is immersed in a \SI{2}{\tesla} axial magnetic field
and provides charged-particle tracking in the range $|\eta| < 2.5$.
A high-granularity silicon pixel detector typically provides four measurements per track and
is surrounded by a silicon microstrip tracker (SCT), which usually provides four three-dimensional measurement points per track.
These silicon detectors are complemented by a transition radiation tracker,
which enables radially extended track reconstruction up to $|\eta| = 2.0$.
 
The MS comprises separate trigger and
precision tracking chambers which measure the deflection of muons in a magnetic field generated by superconducting air-core toroids.
The precision chamber system covers the region $|\eta| < 2.7$ with three layers of monitored drift tubes,
complemented by cathode-strip chambers in the forward region, where the background is highest.
The muon trigger system covers the range $|\eta| < 2.4$ with resistive-plate chambers in the barrel and thin-gap chambers in the endcap regions.
 
A two-level trigger system is used to select interesting events~\cite{TRIG-2016-01}.
The level-1 trigger is implemented in hardware and uses a subset of the muon spectrometer and calorimeter information
to reduce the event rate to around~\SI{100}{\kHz}.
This is followed by a software-based trigger which runs offline reconstruction algorithms and reduces the event rate to approximately~\SI{1}{\kHz}.

\section{Data and simulated event samples}
Data recorded in 2015 with the ATLAS detector at the LHC in proton--proton collisions at a centre-of-mass energy of 13~\TeV{} are used in this analysis. The data set corresponds to an integrated luminosity of 3.2~\ifb{}.
Only events recorded when the detector was fully operational are considered.
 
Simulated MC events are used both to estimate the contamination from background processes in data
and to correct the measured data for detector inefficiency and resolution effects
(Section~\ref{sec:unfolding}).
 
The \Zmm{} signal process was simulated using the next-to-leading-order \Powheg{}~\cite{Alioli:2008gx,Alioli:2010xd} event generator with the
CT10 set of parton distribution functions (PDFs)~\cite{Pumplin:2002vw} and interfaced to the \Pythia{}~8.170 event generator~\cite{Sjostrand:2006za,Sjostrand:2007gs} to simulate the parton shower, hadronization and \UE\ with the CTEQ6L1 PDF set and the \textsc{AZNLO} set of tuned parameters~\cite{STDM-2012-23}.
The latter option tunes the event generator to the \Zpt\ measurement at $\sqrt{s}=7\,\TeV$~\cite{STDM-2012-23}. Hence, it retunes the overall \UE\ activity by adjusting the \Pythia{}~MPI cut-off parameter to the \UE\ activity of the previous measurement~\cite{STDM-2011-42} in the lowest \Zpt\  bin ($0$ to $5\,\GeV$).
\Photos{}~\cite{Golonka:2005pn} was used to simulate final-state electromagnetic radiation.
The \Pythia~generator uses \pt-ordered parton showers and a hadronization model based on the fragmentation of colour strings. Its MPI model interleaves the ISR and FSR emissions with MPI scatters.\\
An alternative signal sample used for cross-checks and systematic uncertainty evaluations was simulated using \Sherpa{}~2.2.0~\cite{Gleisberg:2008ta},
which has an independent implementation of the parton shower, hadronization, \UE\ and FSR.
The \Sherpa\ samples utilize the NNPDF30NNLO PDF set~\cite{Ball:2014uwa} and were generated with the nominal tune set of version 2.2.0.
The \Sherpa{} generator uses leading-order matrix elements with a model for MPI similar to that of \Pythia{}~8 but without interleaving the FSR. It implements a cluster hadronization model similar to that of \HERWIGpp{}.
\Sherpa{}~and \Pythia{}~impose the infrared cut-off for MPI as a smooth function. In contrast, \Herwigpp{} implements it as a step function.
A signal sample produced with the MC generator \HERWIGpp{} \cite{Bahr:2008pv} using the \textsc{UE-EE-5} tune \cite{Gieseke:2012ft} provided by the generator's authors and the corresponding \textsc{CTEQ6L1}~PDF set
is compared with unfolded data in Section~\ref{sec:Results}.
This tuning uses energy extrapolation and was developed to describe the \UE\ and double parton interaction effective cross-section.
\Herwigpp{}~uses, similarly to \Pythia{}, a leading-logarithm parton shower model matched to leading-order matrix element calculations, but it implements a cluster hadronization scheme with parton showering ordered by emission angle.

Three sources of background are estimated using MC samples: \Ztautau{}, $WW\rightarrow\mu\nu\mu\nu$, and the \ttbar{} process,
each of which was simulated using \Powheg{} \cite{Melia2011,Frixione:2007nw} interfaced to \Pythiaeight{} or \Pythiasix{} for \ttbar{}.
The \Pythia{} tune set for \Ztautau{} and $WW\rightarrow\mu\nu\mu\nu$ is the same as was used for the signal process (AZNLO).
The Perugia~2012~\cite{Skands:2010ak} tune set was used for simulation of the \ttbar{} process.
 
Overlaid MC-generated minimum-bias events~\cite{SOFT-2010-01} simulate the effect of multiple interactions in the same bunch crossing (pile-up).
These samples were produced with \Pythia{}~8 using the A2 tune set \cite{ATL-PHYS-PUB-2011-014} in combination with the \textsc{MSTW2008LO} PDF set. The A2 tune set was matched to the ATLAS minimum-bias measurement at $\sqrt{s}=\SI{7}{\tera\electronvolt}$ \cite{STDM-2011-14}.
The mean number of interactions per bunch crossing $\langle\mu\rangle$ during the 2015 data-taking with \SI{25}{\nano\second} bunch spacing was 13.5.
The simulated samples are reweighted to reproduce the distribution of the number of interactions per bunch crossing observed in the data.\\
 
The \Geant4~\cite{Agostinelli:2002hh} program simulated the passage of particles through the ATLAS detector.
Differences in muon reconstruction, trigger, and isolation efficiencies between MC simulation and data are
evaluated using a tag-and-probe method~\cite{PERF-2015-10}, and the simulation is corrected accordingly.
Additional factors applied to the MC events correct for the description of the muon energy and momentum scales and resolution,
which are determined from fits to the observed \Zboson{} boson line shapes in data and MC simulations~\cite{PERF-2015-10}.
Finally, correction factors adjust the distribution of the longitudinal position of the primary $pp$ collision vertex~\cite{ATLAS-CONF-2010-069} to the one observed in the data.
 
\section{Event and track selection}
Candidate \Zmm{} events are selected by requiring that at least one out of two single-muon triggers be satisfied.
A high-threshold trigger requires a muon to have $\pT{} > 40~\GeV$, whilst a low-threshold trigger requires $\pt > 20~\GeV$~and the muon to be isolated from additional nearby tracks.
All events are required to have a primary vertex (PV). The PV is defined as the reconstructed vertex in the event with the highest $\Sigma \pt$ of the associated tracks, consistent with the beam-spot position (spatial region inside the detector where collisions take place) and with at least two associated tracks with $\pt > 400\,\MeV$.
 
The main selections to define the regions of phase space are summarized in Table \ref{tab:MeasurementDefinition}.
The reconstruction procedure for muon candidates combines tracks reconstructed in the inner detector with tracks reconstructed in the MS~\cite{PERF-2015-10}.
The reconstructed muons are required to have \pt $>$ 25~\GeV\ and $|\eta|<2.4$.
Track quality requirements are imposed to suppress backgrounds,
and the muon candidate is required to be isolated using a \pt{}- and $\eta$-dependent `gradient' isolation criterion~\cite{PERF-2015-10} based on track and calorimeter information.
Muon candidates consistent with having originated from the decay of a heavy quark are rejected by requiring the significance of the transverse impact parameter ($|d_{0}/\sigma(d_{0})|$, with $d_{0}$ representing the transverse impact parameter and $\sigma(d_{0})$ the related uncertainty) to be below $3$.
Furthermore, the muon candidates must be associated to the PV, i.e.\ the longitudinal ($|z_{0}\sin{\theta}|$) impact parameter is less than \SI{0.5}{\milli\metre}. The variables $d_{0}$ and $z_{0}$ are measured relative to the PV.
 
Events are required to have exactly two opposite-charged muons satisfying the selection criteria above.
The invariant mass of the dimuon system must be between 66~\GeV{} and 116~\GeV{}.
 
Tracks reconstructed in the ID from the passage of charged particles are used to form the \UE\ observables.
Each reconstructed track is required to have $\pt> 0.5~\GeV$, $|\eta|<2.5$,
one hit in the innermost layer is required (if expected) and in total at least one hit in the pixel detector and at least six hits in the SCT.
The tracks must have been assigned to the PV, i.e.\ the transverse and longitudinal impact parameters of the tracks relative to the PV must be smaller than $\SI{2}{\milli\metre}$ and $\SI{1.5}{\milli\metre}$ respectively.
An additional requirement on the quality of the fit of the track to the hits in the detector applies to tracks with $\pt> 10~\GeV$ in order to suppress mismeasured tracks at high \pt{}. This criterion affects mainly the tracks associated with the muon candidates and has little impact on the predominantly low-\pt{} tracks of the \UE\ activity. \\
 
The kinematics of the \Zboson{} boson and of the charged particles in the event define the phase space of the fiducial region (particle level).
This closely reflects the selection made on measured detector quantities outlined before.
Simulated events are required to have two prompt muons that satisfy $\pt>25$~\GeV{} and $|\eta|<2.4$ with each muon defined at the `bare' level (after final-state QED radiation).
The measurements are all reported in bins of \Zpt, the results presented in this paper are not sensitive to the predicted shape of the \Zpt\ spectrum, even though they are sensitive to jet activity in the event.
As a cross-check the observables are constructed as defined before but the muons are unfolded to the `dressed' level (i.e.\ collinear QED FSR is added to the ‘bare’ level muons) similar to the previous \UE\ measurement in \Zboson{} events \cite{STDM-2011-42}. The difference between the results after unfolding to different generator levels is below the percent level and is less than the uncertainty related to the unfolding procedure.
Charged particles must be stable, i.e.\ have a proper lifetime with $c\tau>\SI{10}{\milli\metre}$, with $\pt>0.5$~\GeV{}. and $|\eta|<2.5$.
 
The statistical uncertainties of the data and the MC simulations are propagated using the bootstrap method~\cite{efron1979}.
While the statistical error of the data is the limiting factor for all distributions at high \Zpt, it does not limit the measurements in phase-space regions of lower \Zpt, which are particularly important for tuning MC simulations.
\newenvironment{DIFnomarkup}{}{}
 
\begin{DIFnomarkup}
\begin{table}[thp]
\begin{center}
\caption{A summary of the fiducial volume definition of the measurement, the particle-level definition, and the main observables.
The first row lists selection criteria for the signal muons (indicated with an $\mu$ as superscript) limited by the detector geometry, while the cut on the dimuon invariant mass $m^{\ell\ell}$ yields a low background contamination.}
\label{tab:MeasurementDefinition}
\footnotesize
\begin{tabular*}{\textwidth}{l @{\extracolsep{\fill}} | l l l l }
\hline
\hline
Fiducial volume (for muon selection)                & $p_\mathrm{T}^{\mu}>\SI{25}{\GeV}$, $|\eta|<2.4$,   $\SI{66}{\GeV}<m^{\mu\mu}<\SI{116}{\GeV}$    \\
\hline
Particle-level definition    & $\particlept{}>\SI{0.5}{\GeV}$, $|\eta|<2.5$, charge $\neq 0$, stable (i.e.\ a proper lifetime of $c\tau>\SI{10}{\milli\metre}$) \\
\hline
\hline
\end{tabular*}
 
\end{center}
\end{table}
\end{DIFnomarkup}
 
\section{Corrections and systematic uncertainties}

\subsection{Unfolding}
\label{sec:unfolding}
 
An iterative Bayesian unfolding technique is used to correct the data for detector inefficiencies and resolution~\cite{DAgostini:1994zf,DAgostini:2010,Adye:2011gm}.
Response matrices connect each observable at the detector and particle levels; these are constructed using the \Powheg{}+\Pythiaeight{} signal MC sample which is overlayed with pile-up events at detector level.
Each response matrix corresponds to a bin of \Zpt{} or thrust, with the migration of events between \Zpt{} or thrust bins corrected using a per-bin purity correction factor.
In the context of MC simulations, the purity of one bin is defined as the fraction of events that are reconstructed in the same bin as the original particle level quantity.
The bin intervals in \Zpt{} and thrust are chosen to yield high purities ($>0.9$ for the bins in \Zpt{} and $>0.85$ for the two bins in \thrust{}) enabling the per-bin corrections.
For the observable \ptspec{}, two unfolding iterations are sufficient for convergence of the unfolding results, while for all other observables eight iterations are performed.
The evaluation of the mean value of each observable in a bin of \Zpt{} and thrust occurs after unfolding.
The bin boundaries are the same at both the detector and particle levels.
 
\subsection{Background subtraction}
 
The background contributions to the selected data from the \Ztautau{}, \ttbar{},
and $WW\rightarrow\mu\nu\mu\nu$ processes are estimated using MC simulations.
In total, these are about 0.7\% of selected data events. This fraction varies from 0.9\% for the lowest bin in \Zpt\ to the per mille level for the highest \Zpt\ bin.
The background contribution from multijet processes is estimated using a data-driven technique based on the isolation and charge of the two reconstructed muons, similar to previous analyses~\cite{STDM-2011-42}.
The size of the multijet contribution in the data is less than 0.1\%.
The unfolding of the data is done after the subtraction of all MC and data-driven background estimates.

\subsection{Systematic uncertainties}

Systematic uncertainties can arise due to possible mismodelling of the muon momentum scale or resolution, as well as the reconstruction, identification, and isolation efficiencies.
Furthermore, limited knowledge of the ID material distribution \cite{PERF-2015-07} dominates the uncertainties in the track reconstruction inefficiencies.
Also the effect of falsely reconstructed tracks (when there is no corresponding charged particle) contributes to all observables.
 
All uncertainties related to imperfect modelling of the detector are assessed using MC simulations.
The data are first unfolded using the nominal MC simulation samples.
Then the data are unfolded with MC samples where the parameter of the simulation which is affected by the mismodelling is varied by $\pm1\sigma$ of its estimated uncertainty.
The average of the up and down shifts is assigned as the corresponding systematic uncertainty.
 
Since the observables are primarily track-based, the track-related systematic uncertainties dominate the total detector-related uncertainty.
These are of the order of 2\% regardless of the observable and region.
Systematic uncertainties related to the muon reconstruction are a negligible fraction of the overall uncertainty.\\
 
Uncertainties due to mismodelling of the background processes are also considered.
For the background processes modelled with MC simulations, the electroweak background normalization is varied by $\pm5\%$ and the \ttbar{} background normalization by $\pm15\%$ (approximately within their theoretical uncertainties \cite{STDM-2011-24,TOPQ-2011-01}) and the effect on the final measurements is estimated.
The full effect of including the multijet background or not is taken as an uncertainty.
The combined background-related uncertainties form a negligible fraction of the total systematic uncertainty. The dependence of the background uncertainty on \Zpt{} is negligible for this measurement.
 
An important consideration for these measurements is the modelling of the pile-up, since the MC simulations must correct for contamination from pile-up tracks through the unfolding procedure.
When averaging over all simulated events about 13\% of the selected tracks which are compatible with the primary vertex originate from pile-up.
 
A variation in the pile-up reweighting of the MC simulations is included to cover the uncertainty on the ratio between the predicted and measured inelastic cross-section in the fiducial volume defined by $M_{X} >\SI{13}{\giga\electronvolt}$ where $M_{X}$ is the mass of the hadronic system \cite{STDM-2015-05}.
The value of $\left<\mu\right>$ assumed in the MC simulations for the unfolding process is varied by $\pm9\%$ from the nominal value.
This uncertainty in the pile-up modelling is one of the largest sources of systematic uncertainty in the tails of the distributions of \particlept{}, \nch{}, \sumpt{}, and \meanpt{},
and for the mean distributions.
The uncertainties related to the inaccuracies of the detector and pile-up modelling are combined and referred to as the `Detector' uncertainty in the following figures.
 
Two additional cross-checks validate the pile-up modelling and the consistency of removing the pile-up effects via the unfolding technique.
First, the unfolding procedure for all observables in all measurement bins is repeated for three intervals of $\left<\mu\right> $, namely $[8$--$10]$, $[11$--$13]$ and $[14$--$16]$. A mismodelling of pile-up in MC simulations would manifest itself less in the interval of $8\leq \left<\mu\right>\leq10$ and more in the interval of $14\leq \left<\mu\right> \leq16$. The unfolded results for the three intervals are found to be fully compatible within their associated statistical uncertainties, confirming the consistency of the handling of pile-up in the unfolding process.
 
Secondly, a complementary data-driven technique based on the \textit{Hit Backspace Once More} (HBOM) method ~\cite{Monk:2011pg} is used.
The intention is to reproduce pile-up contaminations as realistically as possible.
Hence, the track information associated with non-primary vertices in the data is bundled to form a pile-up library.
A random sample is drawn from this library and used as an example of pile-up effects in data.
If this random sample is added to an individual event, the pile-up effect increases.
A sampling of the library is subsequently used to pollute events with additional pile-up.
Six iterations of pollution are applied, i.e.\ up to six random  samples from the pile-up library are added to each event.
Then the observables are constructed from these additionally contaminated events.
Assuming the values of the observables evolve smoothly with each iteration of additional pile-up, an extrapolation in each bin to the value with zero pile-up vertices yields the HBOM estimate of pile-up subtracted data.
The data are subsequently unfolded using a version of the \Powheg{}+\Pythia{} signal MC samples without pile-up vertices.
The results obtained using this method are consistent with the nominal procedure, and no additional uncertainty is assigned.
 
The uncertainty associated with the unfolding technique is evaluated using a data-driven method.
It accounts for the dependence of the unfolding on the usage of prior knowledge from the MC simulation, i.e.\ the particle level quantities.
The ratio of data to simulation at detector-level is evaluated and smoothed for each observable.
The smoothed ratio is then used to reweight the simulations by applying the event-weight according to the particle level quantity.
The reweighted detector-level distribution is then unfolded using the regular response matrix.
The relative difference between the reweighted particle-level distribution and the reweighted and unfolded detector-level distribution is treated as a systematic uncertainty.
This dependence on prior knowledge from the MC simulation is the dominant systematic uncertainty in most distributions at lower values of \Zpt{}.
An additional method of estimating the uncertainty related to the unfolding is  to unfold the detector-level MC distributions generated with \Sherpa{} using the unfolding matrices based on the \Powheg{}+\Pythia{} MC sample.
The results are compared with the particle level quantities predicted by \Sherpa{}. After taking the uncertainty due to the MC prior into account, a slight discrepancy between the unfolded \Sherpa{} sample and the particle-level distributions remains.
Therefore, an additional contribution to the MC prior uncertainty is introduced to cover this remaining non-closure of the unfolded result and the \Sherpa{} generator level.
In general, it does not exceed the 2--4\% level and is smoothed over the full range of the observable.
In a few cases, this non-closure component dominates the MC prior uncertainty.
These two separate unfolding uncertainties are added in quadrature in all figures.
 
All sources of systematic uncertainty are considered uncorrelated and  are combined in quadrature.
The MC prior uncertainty is one of the largest contributors to the total systematic uncertainty at all values of \particlept{}
and in each \Zpt{} region.
The statistical uncertainty of the data rises with increasing \Zpt{}, contributing a significant fraction of the overall uncertainty.
The breakdown of the individual sources of uncertainties for the four observables, \particlept{}, \nch{}, \sumpt{}, and \meanpt{} is illustrated in Figure \ref{fig:systematics-individual-ptspec} for the example of events with 10 $<$ \Zpt{} $<$ 20~\GeV{} in the trans-min region (the region most sensitive to the \UE), inclusively in $\thrust$.

Figure~\ref{fig:systematics-mean} shows the systematic uncertainties in the arithmetic mean of the \nch{} and \sumpt{} spectra in the trans-min region as a function of \Zpt{} inclusively in $\thrust$.
The largest contributions to the total systematic uncertainties of the mean distributions at all  \Zpt{} values come from either the MC prior uncertainty or the track-related uncertainties.
The statistical uncertainties of the data become large for \Zpt{} greater than around 200~\GeV{}.

\begin{figure}[ht]
\centering
\includegraphics[width=0.49\textwidth]{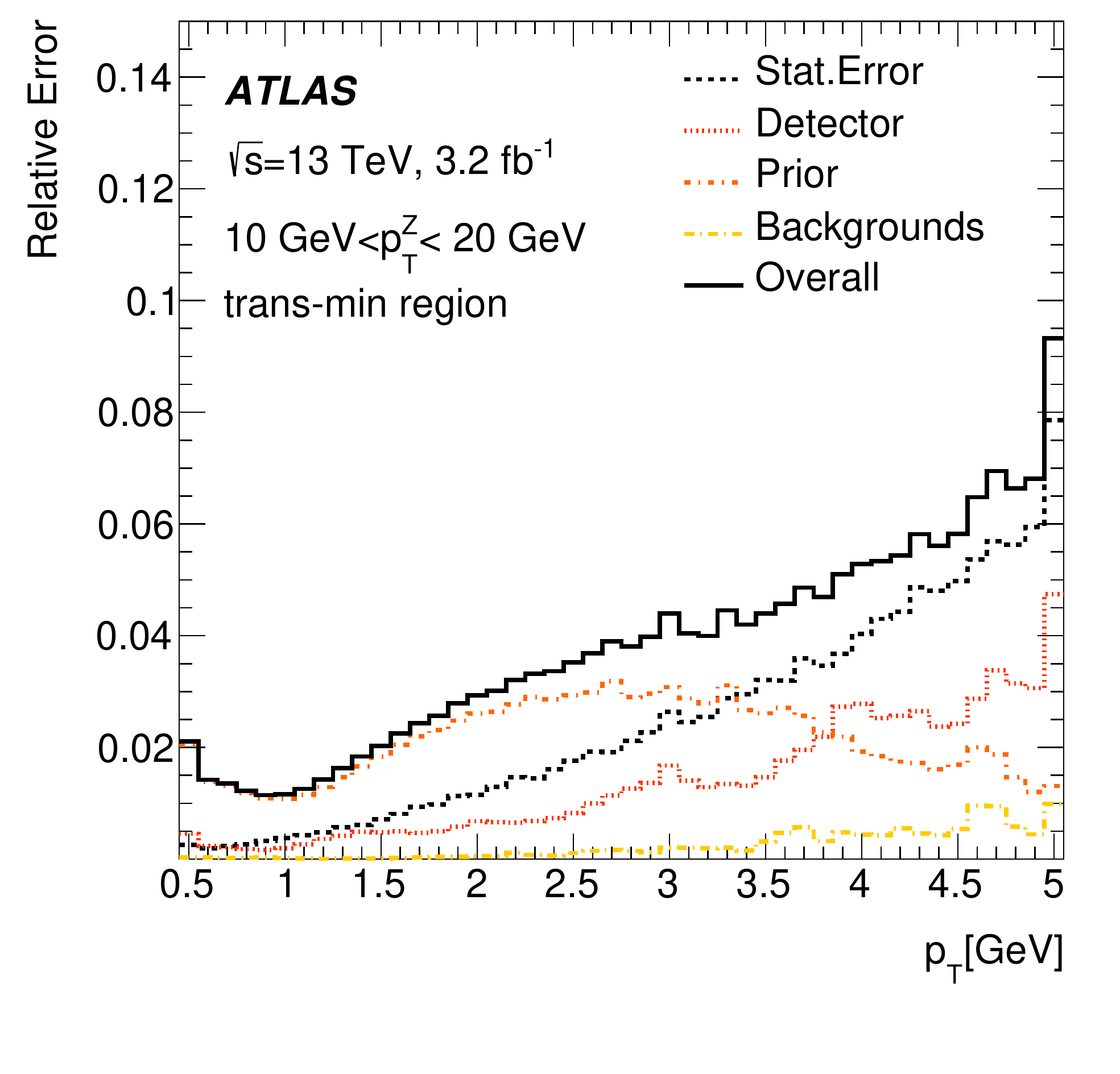}
\includegraphics[width=0.49\textwidth]{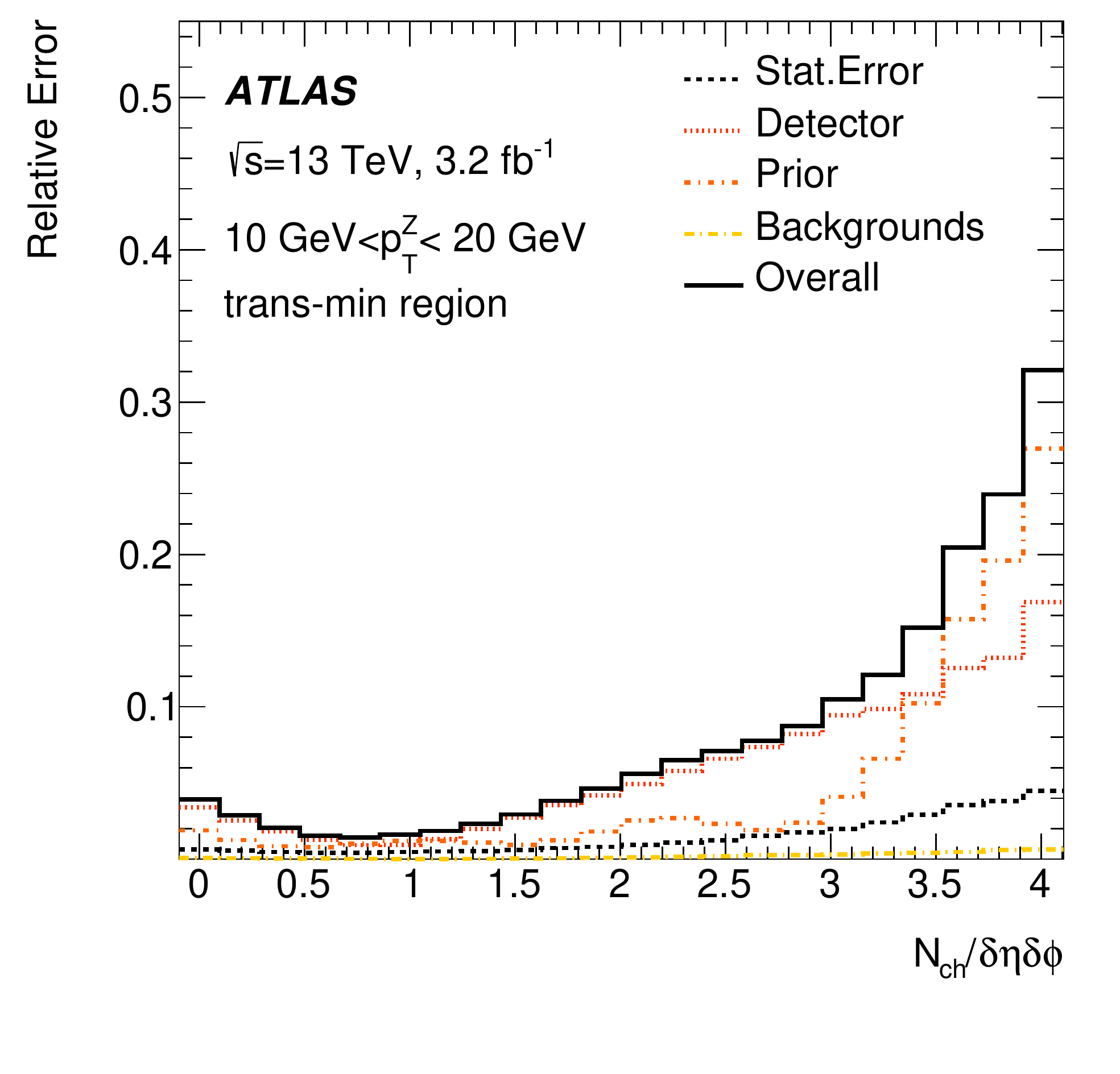}
\includegraphics[width=0.49\textwidth]{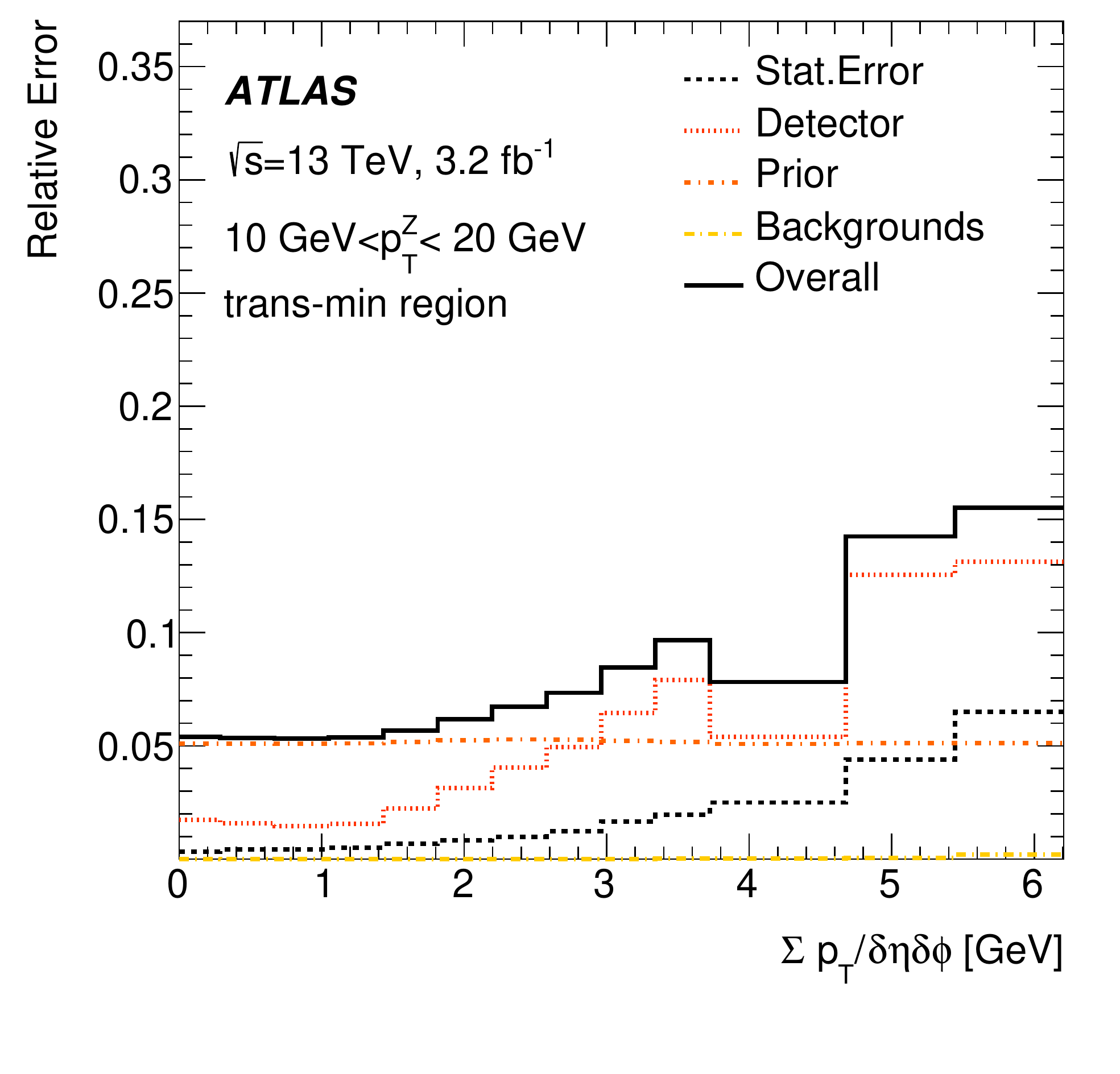}
\includegraphics[width=0.49\textwidth]{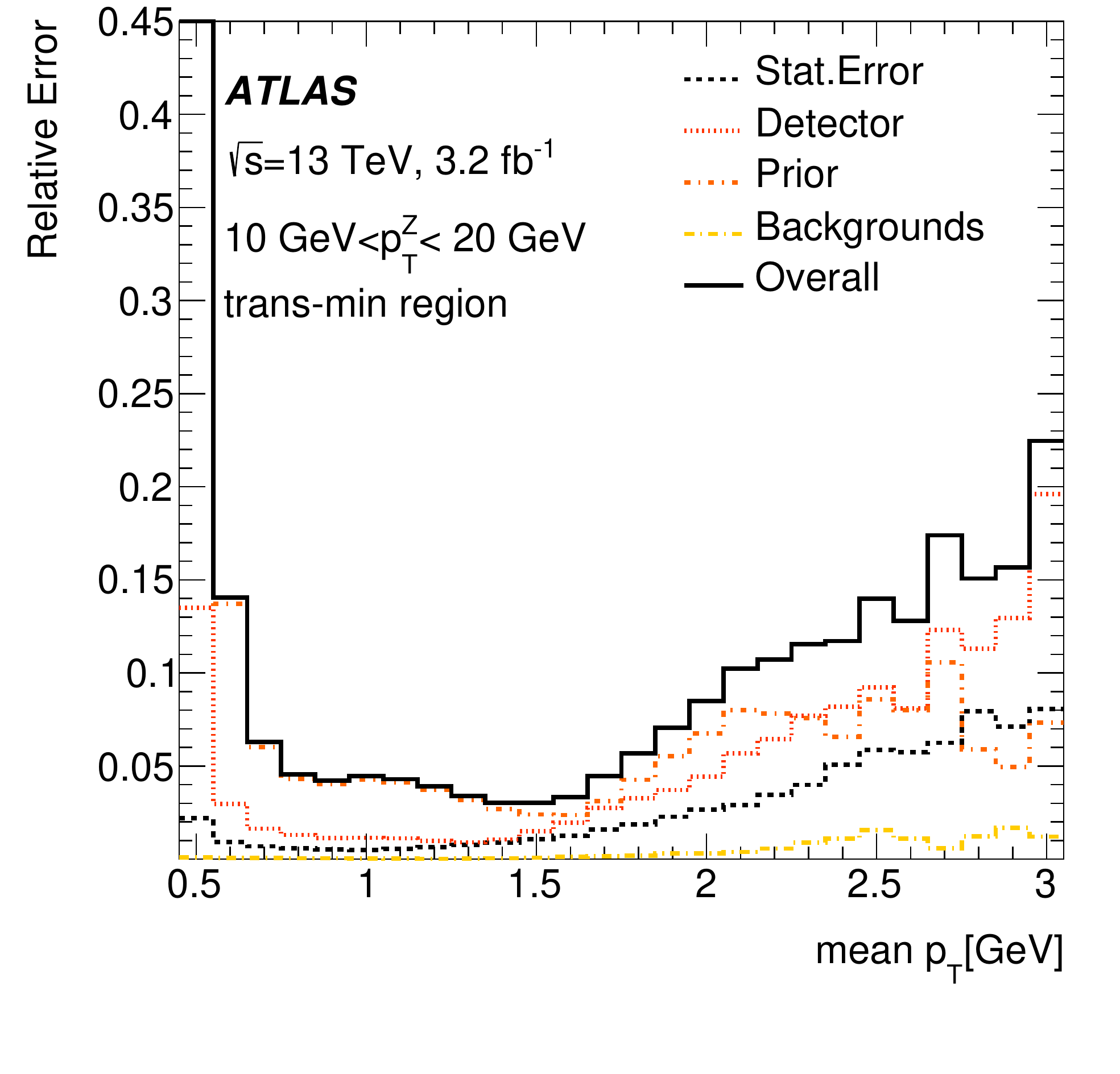}
\caption{Breakdown of systematic uncertainties in the \particlept{} spectrum (upper left), the charged-particle multiplicity (\nch{}, upper right), the scalar sum of the transverse momenta (\sumpt{}, lower left) and the mean transverse momentum (\meanpt{}, lower right) for events with 10 $<$ \Zpt{} $<$ 20~\GeV{} in the trans-min region inclusively in $\thrust$. Here `Prior' combines the two approaches to estimate the unfolding-related uncertainties. `Detector' includes the modelling of the detector and the pile-up conditions.}
\label{fig:systematics-individual-ptspec}
\end{figure}
\begin{figure}[ht]
\centering
\includegraphics[width=0.49\textwidth]{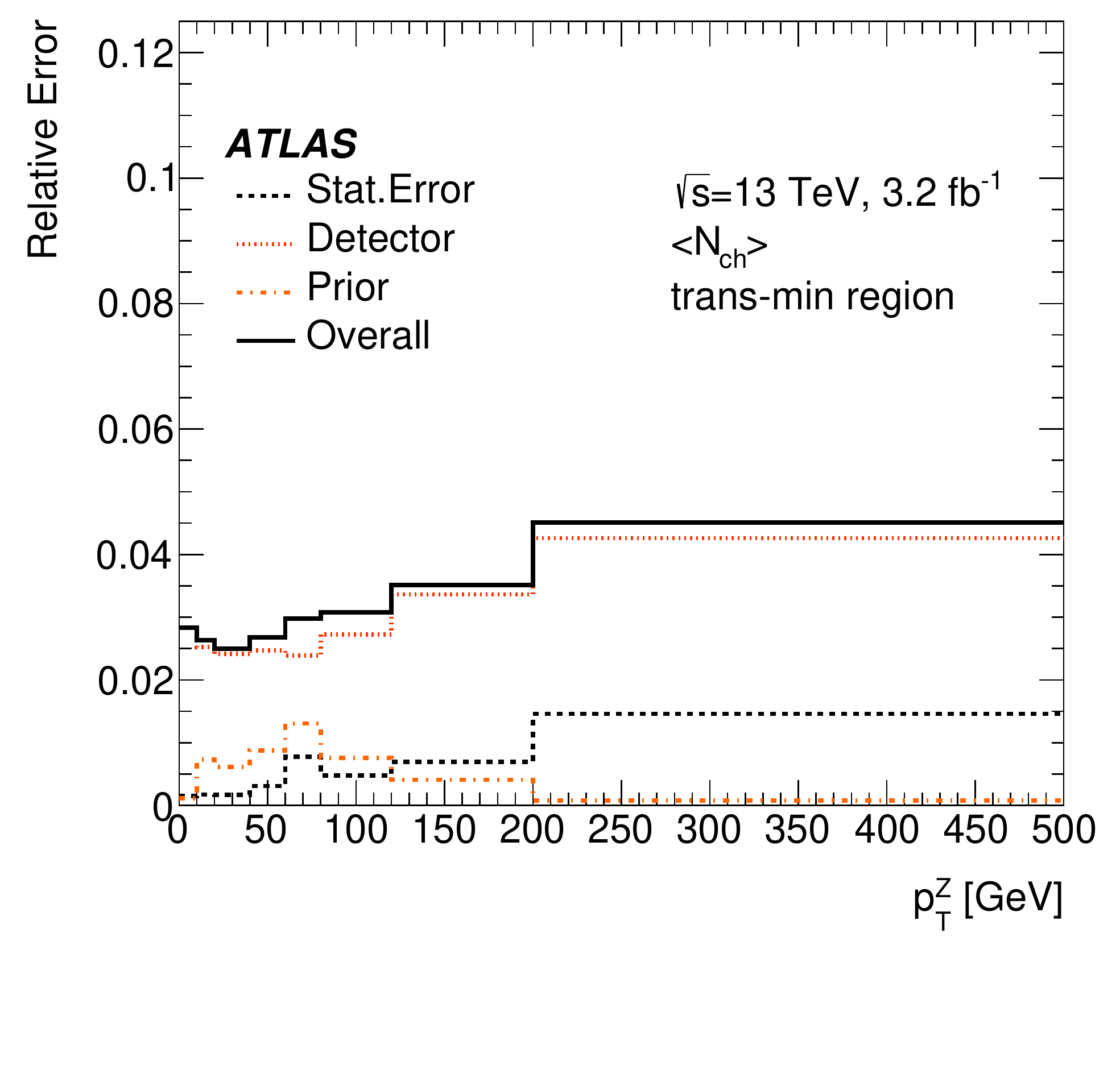}
\includegraphics[width=0.49\textwidth]{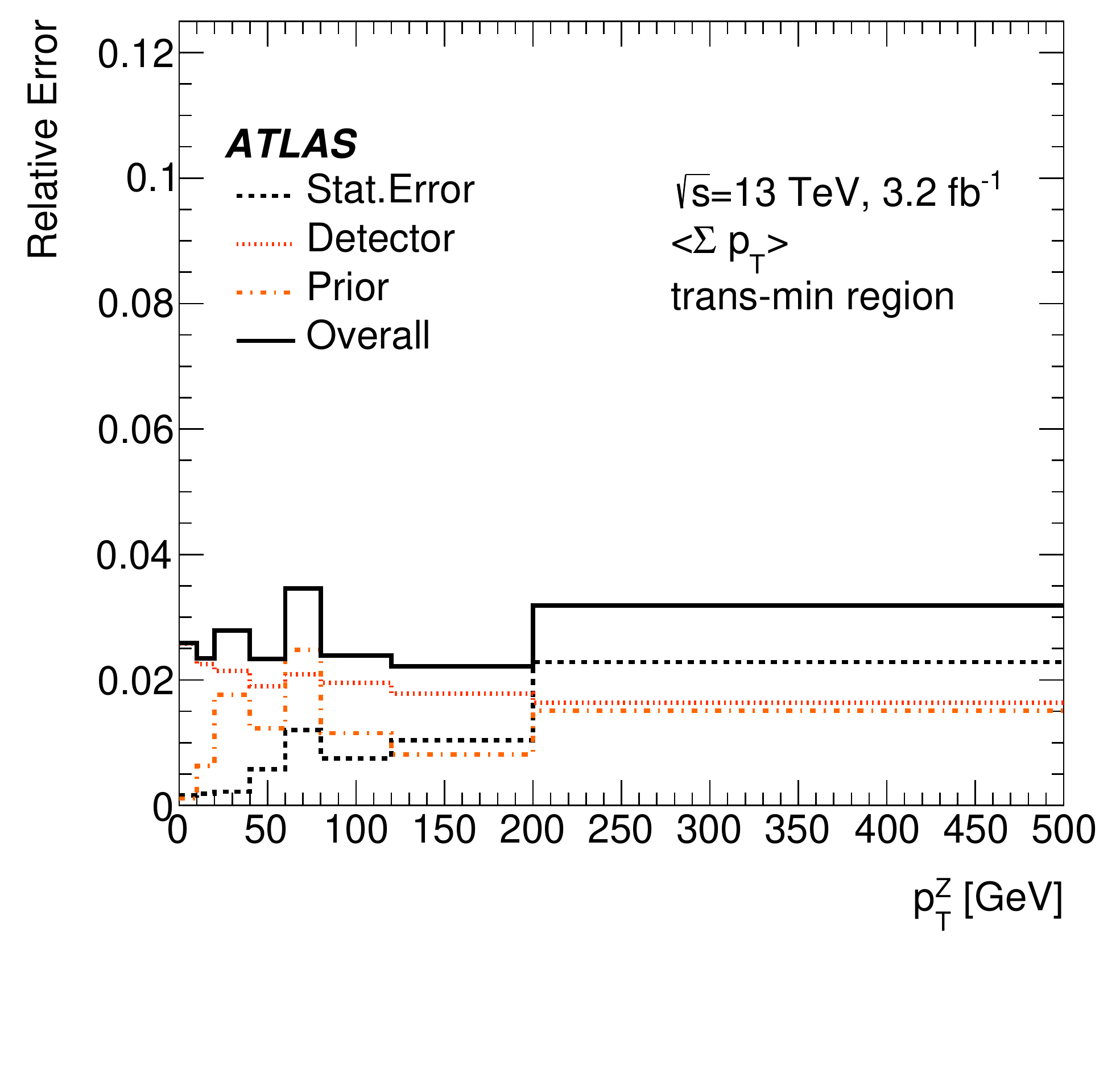}
\caption{A summary of the systematic uncertainties in the arithmetic mean of the \nch{} and \sumpt{} spectra in the trans-min region as a function of \Zpt{}. Here `Prior' combines the two approaches to estimate the unfolding-related uncertainties. `Detector' includes the modelling of the detector and the pile-up conditions.
}
\label{fig:systematics-mean}
\end{figure}
\FloatBarrier
 
\section{\label{sec:Results}Unfolded observables and comparison with model predictions}
 
\subsection{Overview of the results}
 
Distributions of \particlept{}, \nch{}, \sumpt{}, and \meanpt{} are obtained in slices of \Zpt{}
for the different regions defined in the transverse plane and different regions of $\thrust$.
The results for \nch{} and \sumpt{} are normalized relative to the area of the region in $\eta$ and $\phi$.
In addition to the measurements in slices of \Zpt, the arithmetic means of \nch{}, \sumpt{}, and \meanpt{} (\Mnch, \Msumpt{}, and \Mmeanpt) are measured as a function of \Zpt{}.
Only a selection of the most relevant results is discussed in this section: the comparison of the unfolded data to the predictions of different MC generators focuses on the trans-min region.
While the toward region provides insights of similar importance for tuning MC generators after having removed the two muons,
the discussion focuses on the trans-min region to better facilitate comparison with previous measurements.
The \UE\ activity in the toward region is higher compared with that in trans-min. This is expected since the trans-min region is defined as the subregion of the transverse region with the lower activity and for \Zmm events the \UE\ activity is expected to be of similar magnitude in the toward and transverse regions. The trans-min region is statistically less affected by radiation and it is essentially the region where the contribution from ISR is subtracted. Apart from this difference in the amount of activity, the predictive performance of the different MC generators is comparable in the toward and trans-min regions.
No significant difference in the predictive power between these regions is observed.
Both \Mnch{} and \Msumpt{} measured in the trans-min are compared with previous measurements of the \UE\ in \Zboson{} boson events at lower centre-of-mass energies.
 
\subsection{Differential distributions}

Figures~\ref{fig:results-individual-1} and \ref{fig:results-individual-2} show the unfolded \particlept{} spectrum, \nch{}, \sumpt{}, and \meanpt{} for the trans-min region inclusively in $\thrust$ for events with \Zpt{} between 10 and 20~\GeV{} and between 120 and 200~\GeV{}. The predictions from \Powheg{}+\Pythia{}, \Sherpa{}, and \HERWIGpp{} are compared with the data. The ratio of prediction to data is shown beneath each plot.
None of the tested MC generators describes  all aspects of the data well and in some regions the differences exceed the 70\% level.
Generally, the MC generators predict a higher number of particles with small \particlept{} than is observed in data (see top left of Figures~\ref{fig:results-individual-1} and \ref{fig:results-individual-2}).
This is consistent with the MC predictions tending to lower values of \meanpt{}, as is shown on the lower right plots of Figures~\ref{fig:results-individual-1} and \ref{fig:results-individual-2}.
The largest differences  between data and simulation are at low \nch{} and low \sumpt{}, and arise due to the steeper transverse momentum spectrum of charged particles in MC simulations.
\Powheg+\Pythia{} and \Sherpa{} predict a higher fraction of events with fewer charged particles and a consistently smaller sum of \particlept{}. However, \HERWIGpp{} slightly overestimates the fraction of particles with \particlept{} $>$ 2.5~\GeV{} and is qualitatively closer to the shape of the distributions of \nch{} and \sumpt.
With rising \Zpt{}, the data \particlept{} spectrum becomes harder, and \nch{}, \sumpt{}, and \meanpt{} increase. The relative discrepancy remains the same in comparisons with the generator predictions.
 
\begin{figure}[ht]
\centering
\includegraphics[width=0.49\textwidth]{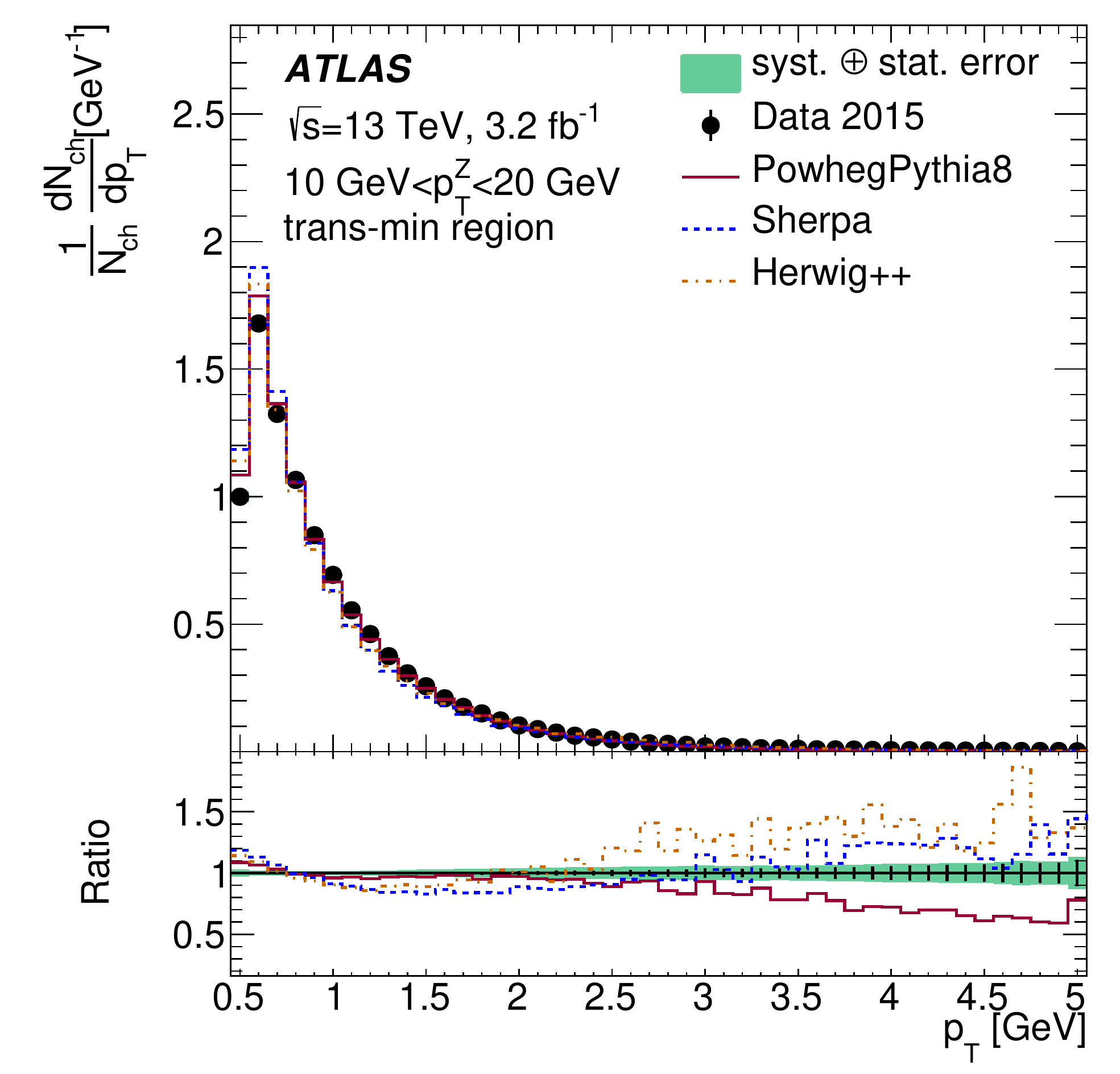}
\includegraphics[width=0.49\textwidth]{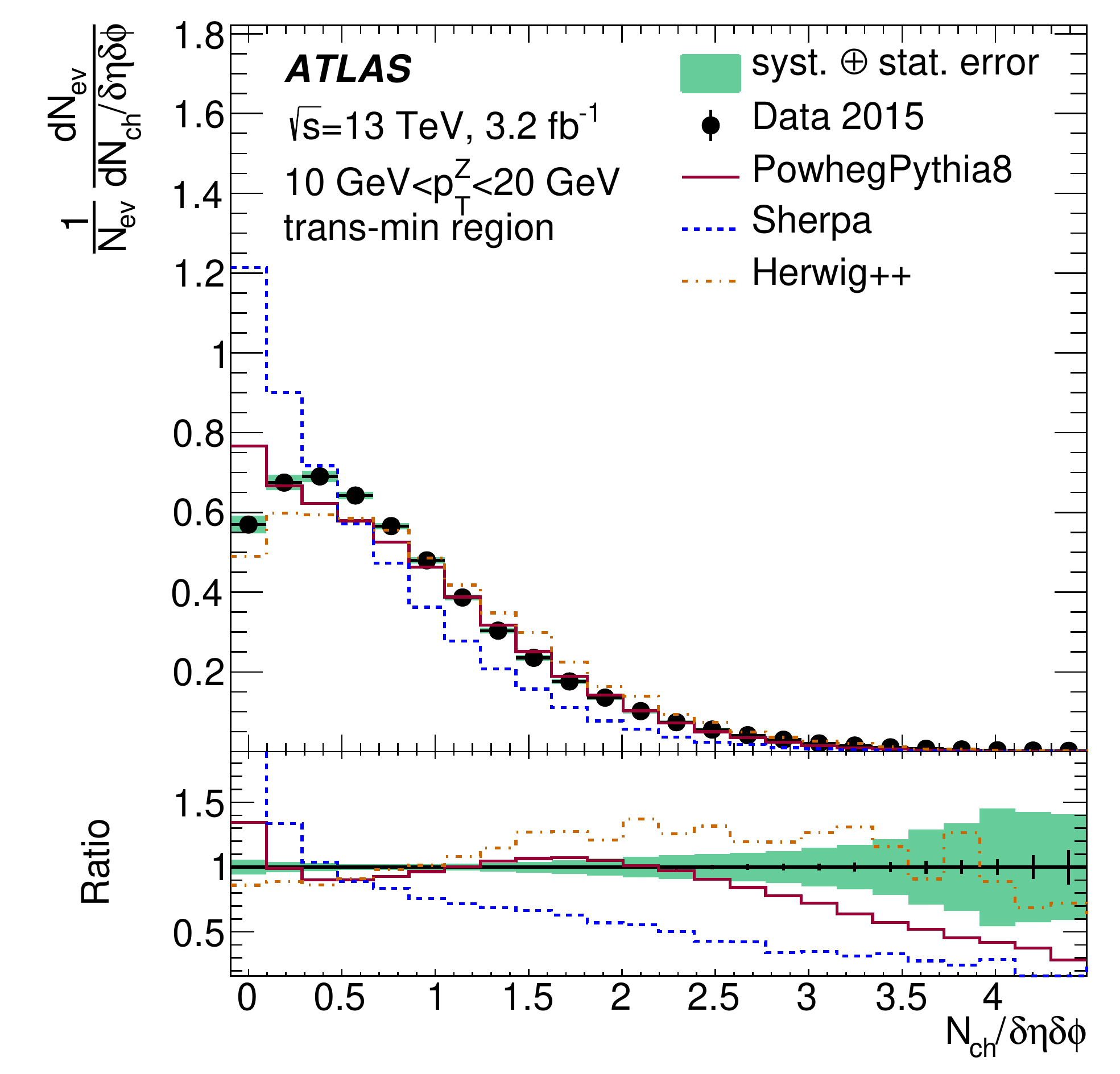}
\includegraphics[width=0.49\textwidth]{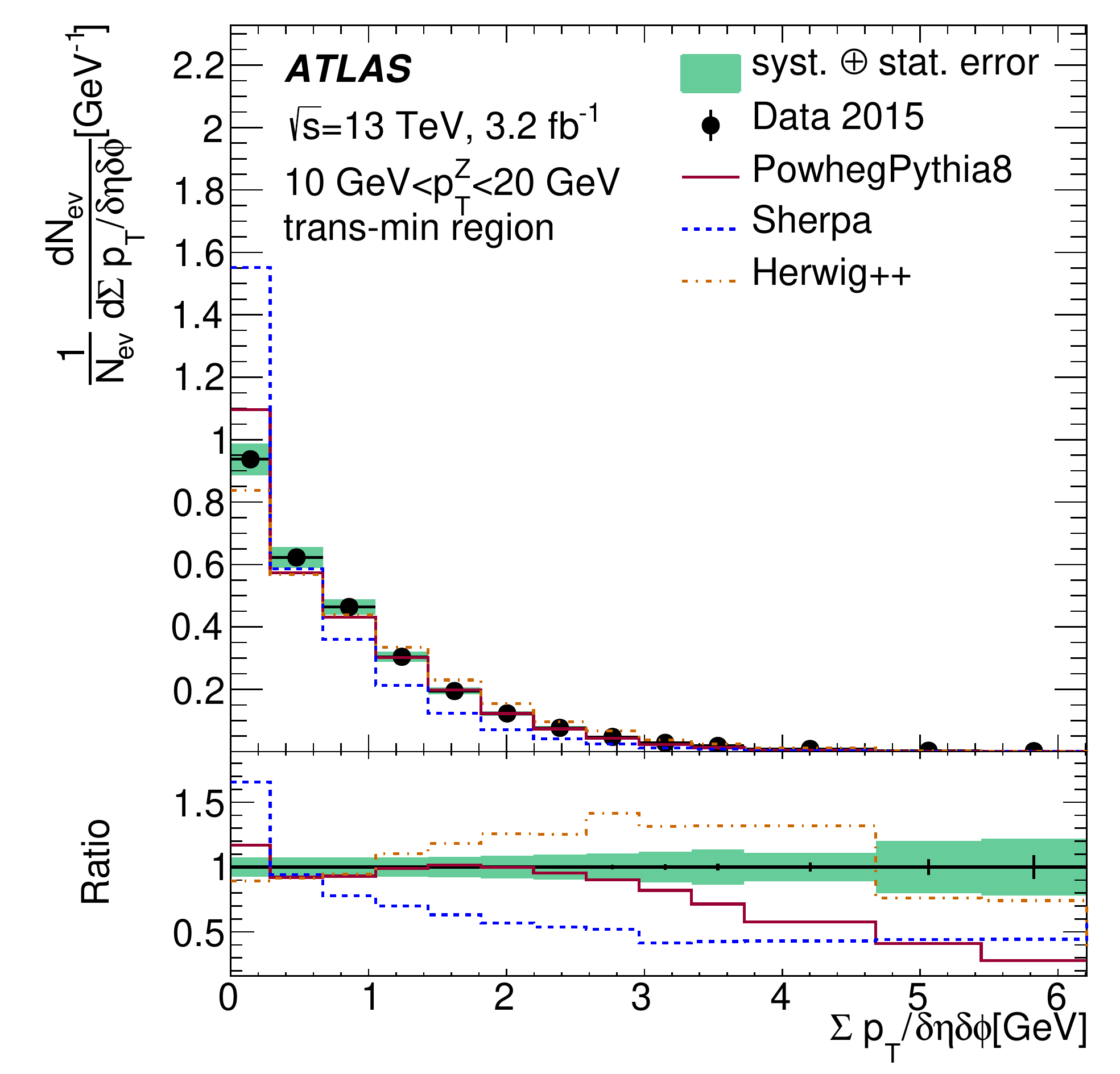}
\includegraphics[width=0.49\textwidth]{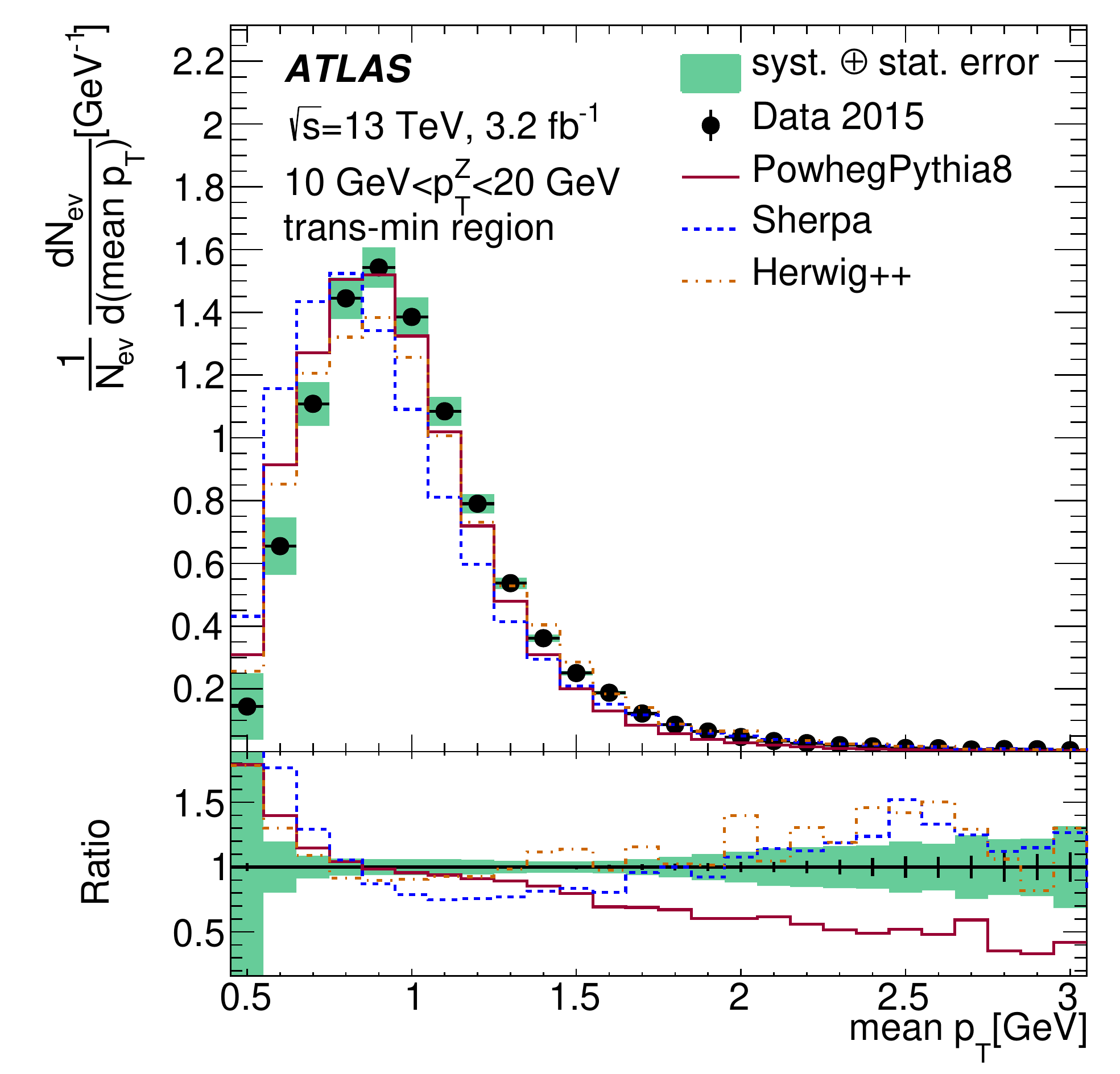}
\caption{Measured spectra of \particlept{} (upper left), the charged-particle multiplicity, \nch{} (upper right), the scalar sum of the transverse momentum of those particles, \sumpt{}, (lower left) and the mean transverse momentum, \meanpt{} (lower right) in the trans-min region inclusively in $\thrust$ for events with 10 $<$ \Zpt{} $<$ 20~\GeV{}. Predictions of \Powheg{}+\Pythia{}, \Sherpa{}. and \HERWIGpp{} are compared with the data. The ratios shown are predictions over data.}
\label{fig:results-individual-1}
\end{figure}
 
\begin{figure}[ht]
\centering
\includegraphics[width=0.49\textwidth]{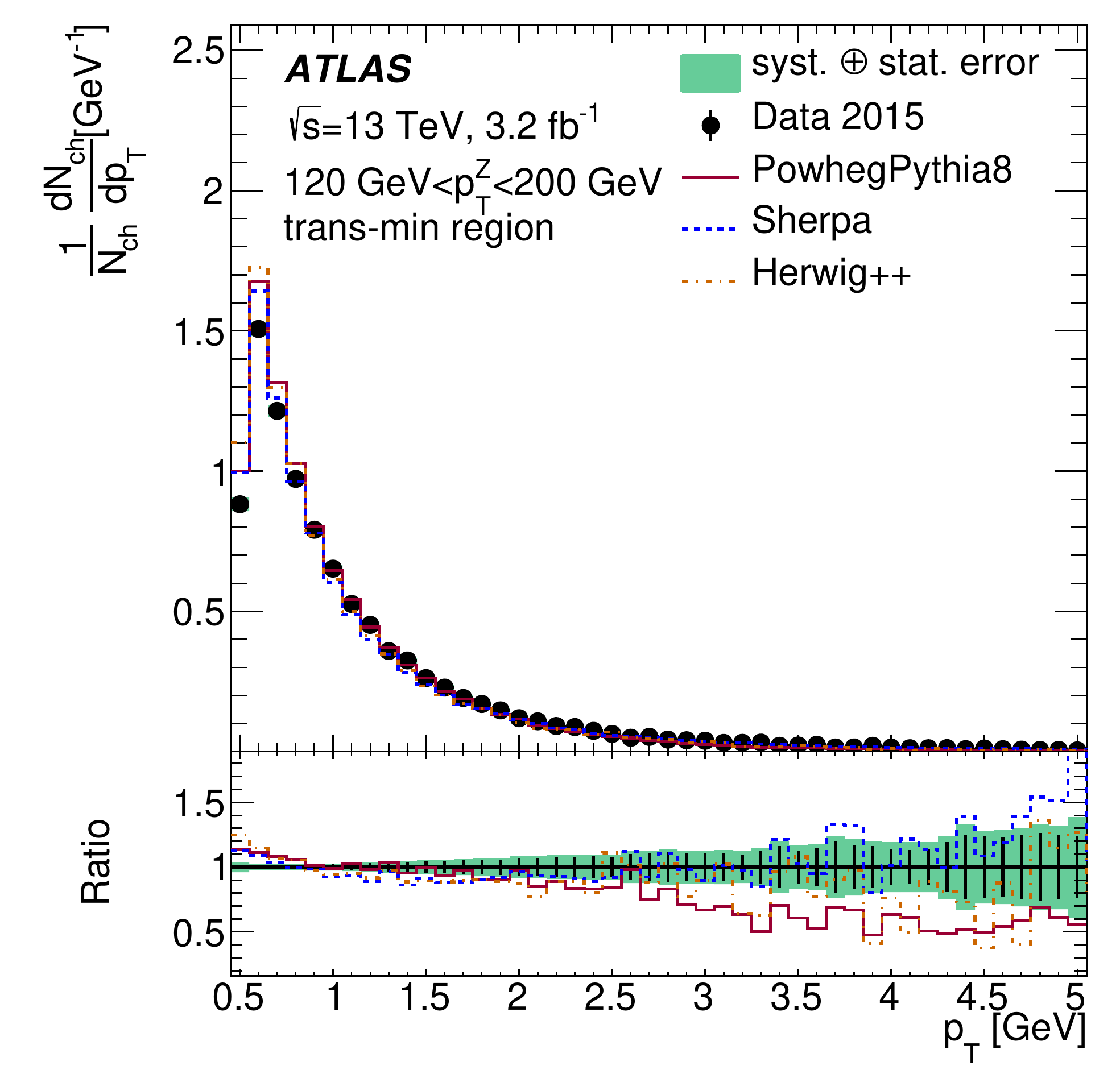}
\includegraphics[width=0.49\textwidth]{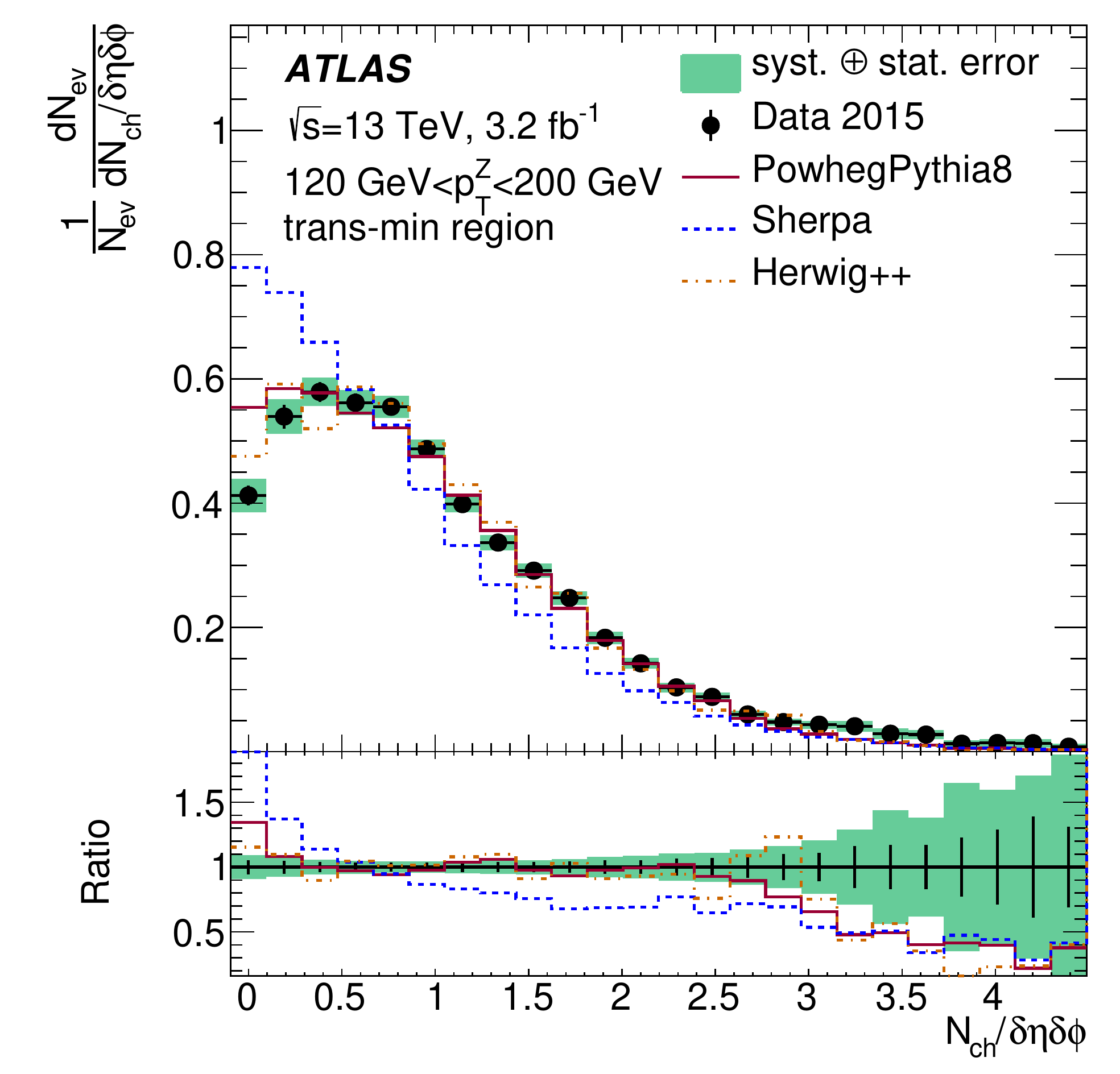}
\includegraphics[width=0.49\textwidth]{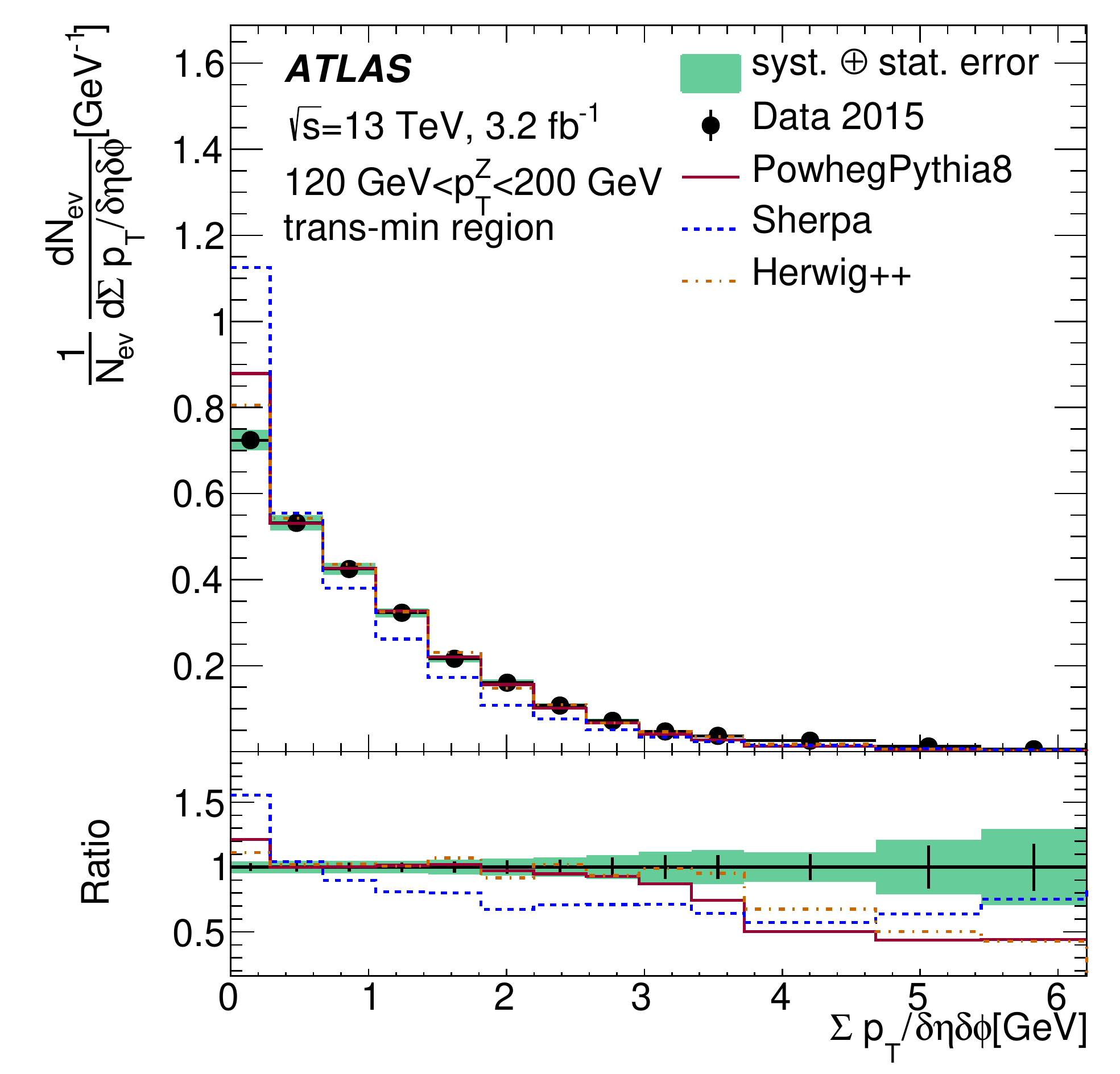}
\includegraphics[width=0.49\textwidth]{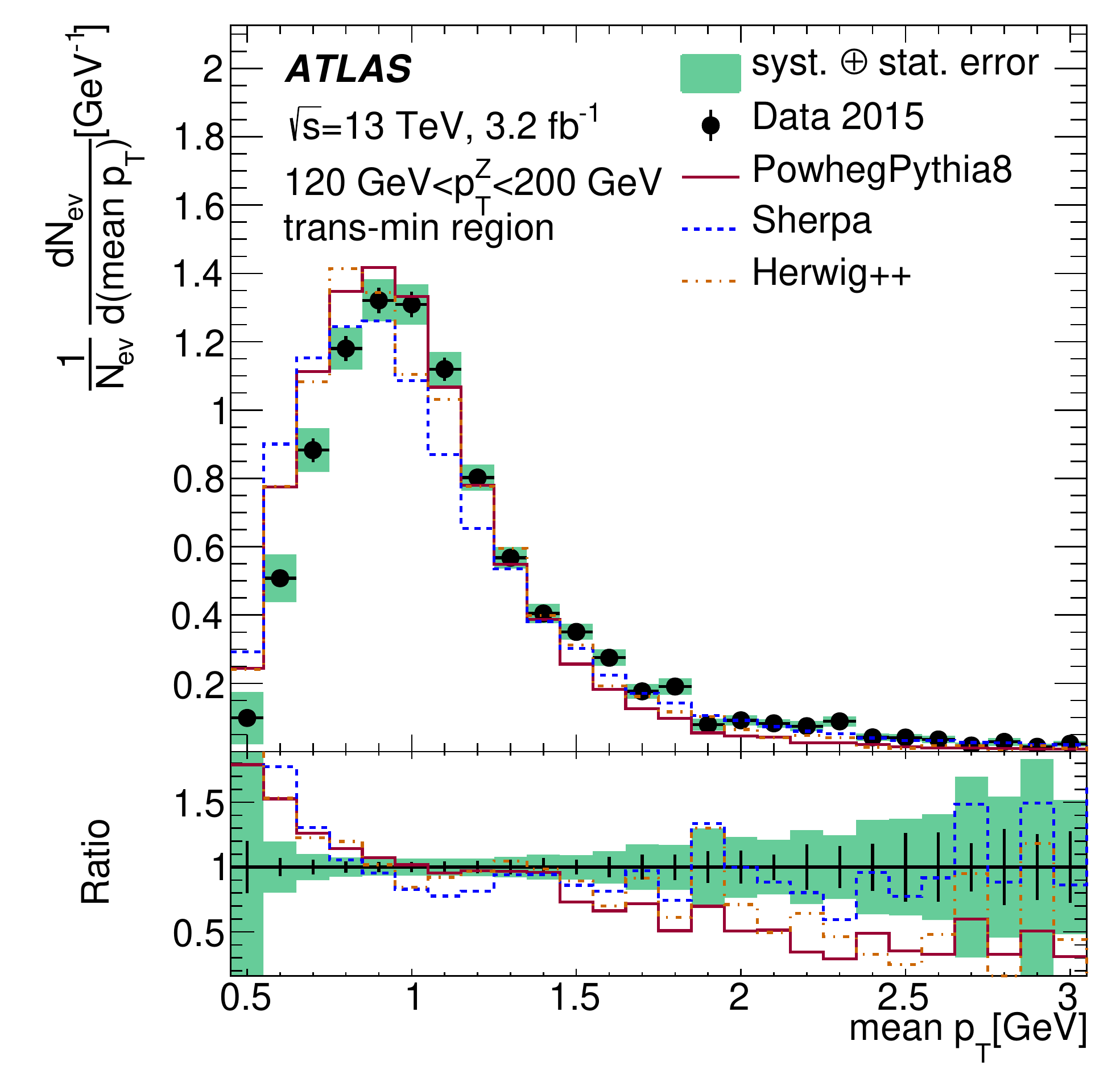}
\caption{Measured \particlept{} spectra (upper left), the charged-particle multiplicity \nch{} (upper right), the scalar sum of the transverse momentum of those particles \sumpt{} (lower left), and the mean transverse momentum, \meanpt{} (lower right) in the trans-min region inclusively in $\thrust$ for events with 120 $<$ \Zpt{} $<$ 200~\GeV{}. Predictions of \Powheg{}+\Pythia{}, \Sherpa{}, and \HERWIGpp{} are compared with the data. The ratios shown are predictions over data.}
\label{fig:results-individual-2}
\end{figure}
 
The dependence on $\thrust$ is illustrated in Figure \ref{fig:results-individual-thrust} for the unfolded \particlept{} spectrum in the trans-min region for events with 10 $<$ \Zpt{} $<$ 20~\GeV{} and 120 $<$ \Zpt{} $<$ 200~\GeV{}.
Similar to the results for the measurement inclusive in \thrust{}, the MC generators predict a higher fraction of particles with low \particlept{} than present in data.
The predictions of \Powheg{}+\Pythia{} are closer to the measured distributions in the lower \Zpt{} region, but \Sherpa{} describes better the full \particlept{} range in the higher \Zpt{} bin. The \Herwigpp{} simulations have significant statistical fluctuations at higher \particlept{}.
The most striking difference between the different regions in \thrust{} is observed for the \Powheg+\Pythia{} generator when focusing on the low \Zpt{} bins for \nch{} as presented in Figure \ref{fig:results-individual-thrust-nTracks}.
In MPI-sensitive regions (left plot in Figure~\ref{fig:results-individual-thrust-nTracks}) the distribution of \nch{} by \Powheg+\Pythia{} is shifted towards higher numbers of charged-particles relative to the data, i.e.\ overshooting the data in the range $1\leq\nch/\delta\eta\delta\phi\leq2.5$.
But in the high thrust region (right plot) the MC generator underestimates the data almost over the full range except for the first two bins.
In contrast, the performances of \Sherpa{} and \Herwigpp{} are consistent when comparing the low and high thrust regions for \nch; \Herwigpp overestimates \nch{}, and \Sherpa{} underestimates it.
The same effect is observed for the distributions of \sumpt\ but is less significant and therefore not presented.
As pointed out in Ref.~\cite{Kar:2018uno}, the regions of high values of \thrust\ are dominated by extra jet activity which is not adequately modelled in \Powheg+\Pythia{}, as shown in the right plots in Figures \ref{fig:results-individual-thrust} and \ref{fig:results-individual-thrust-nTracks}.
 
\begin{figure}[ht]
\centering
\includegraphics[width=0.49\textwidth]{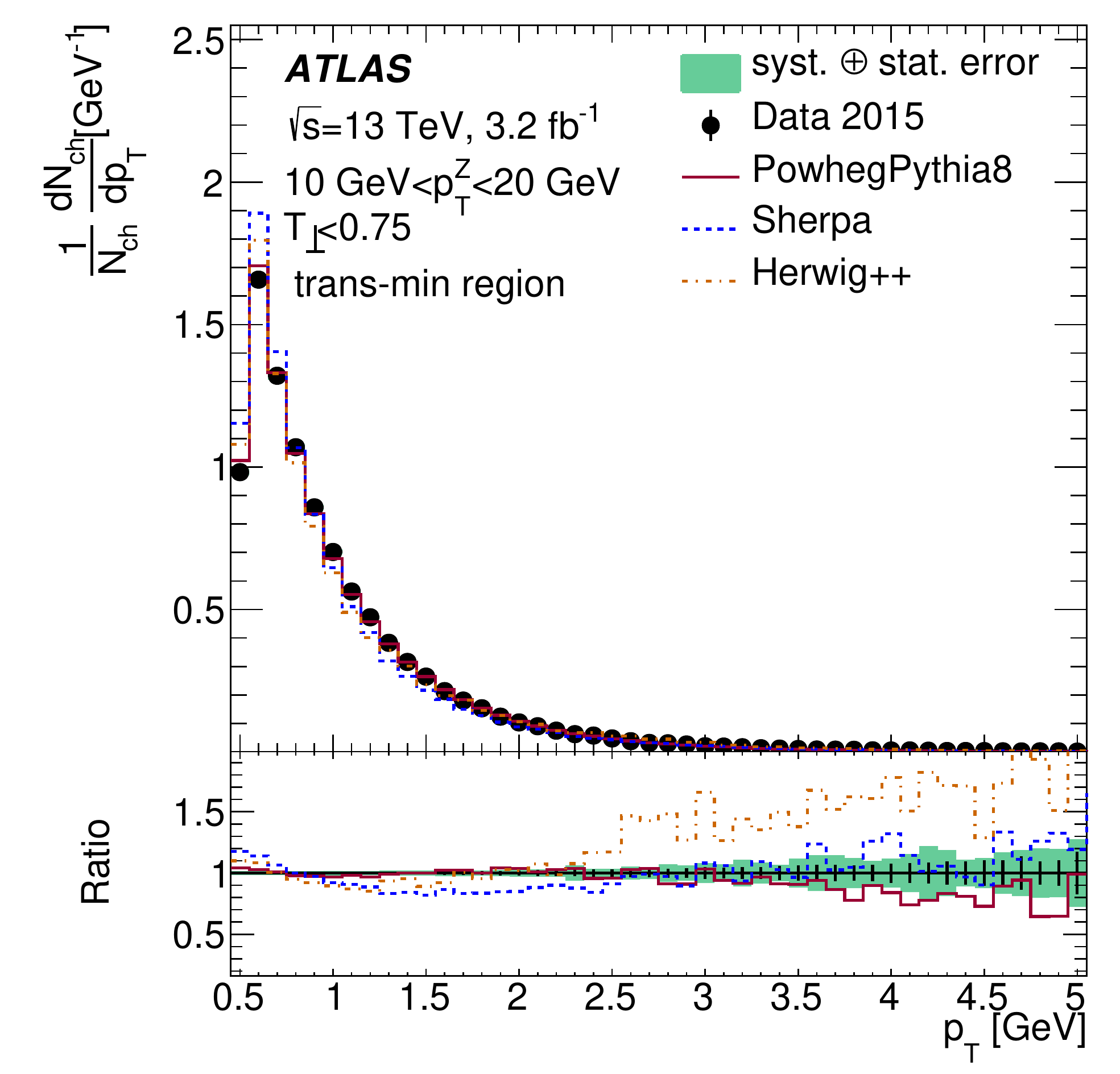}
\includegraphics[width=0.49\textwidth]{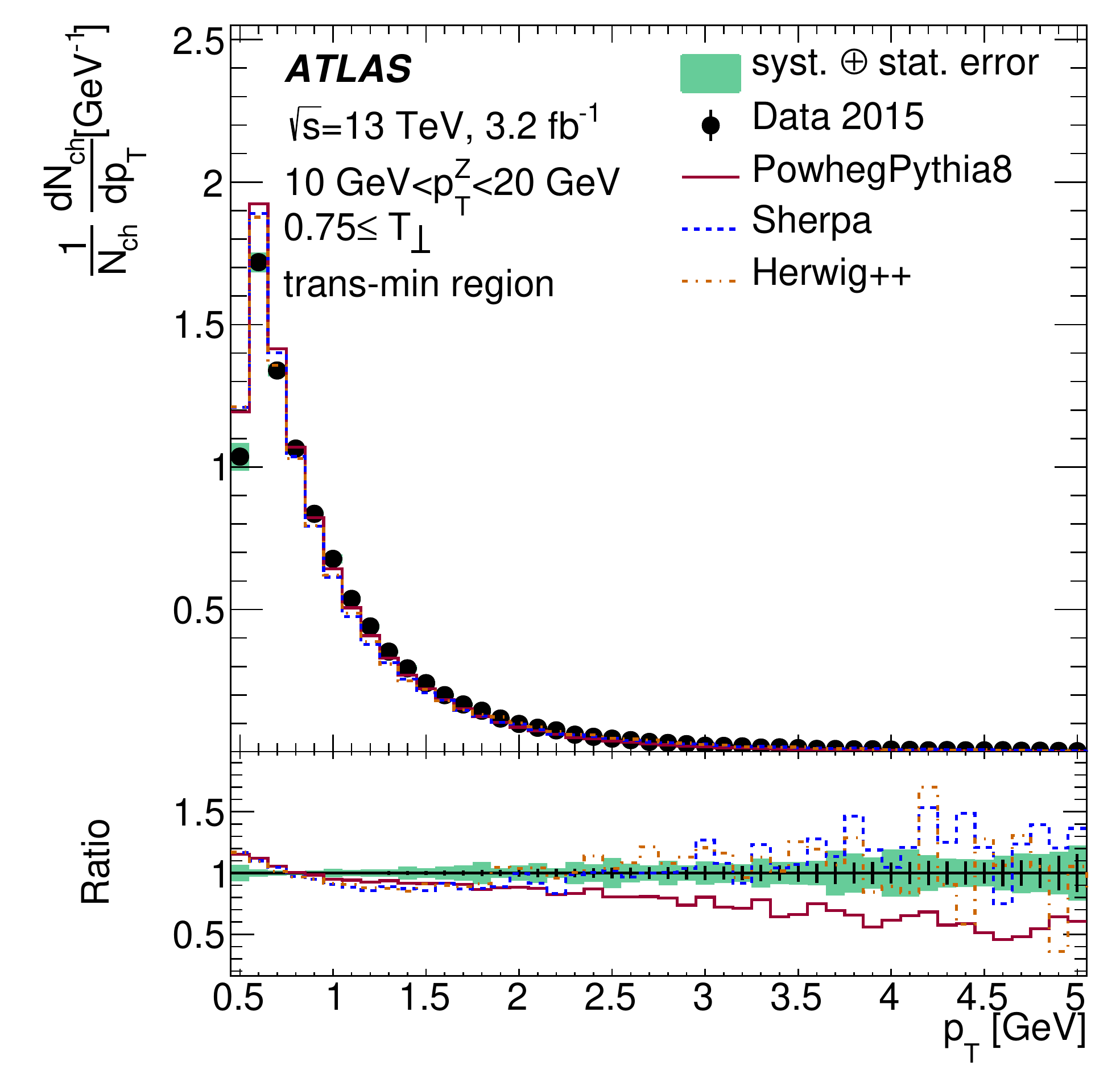}
\includegraphics[width=0.49\textwidth]{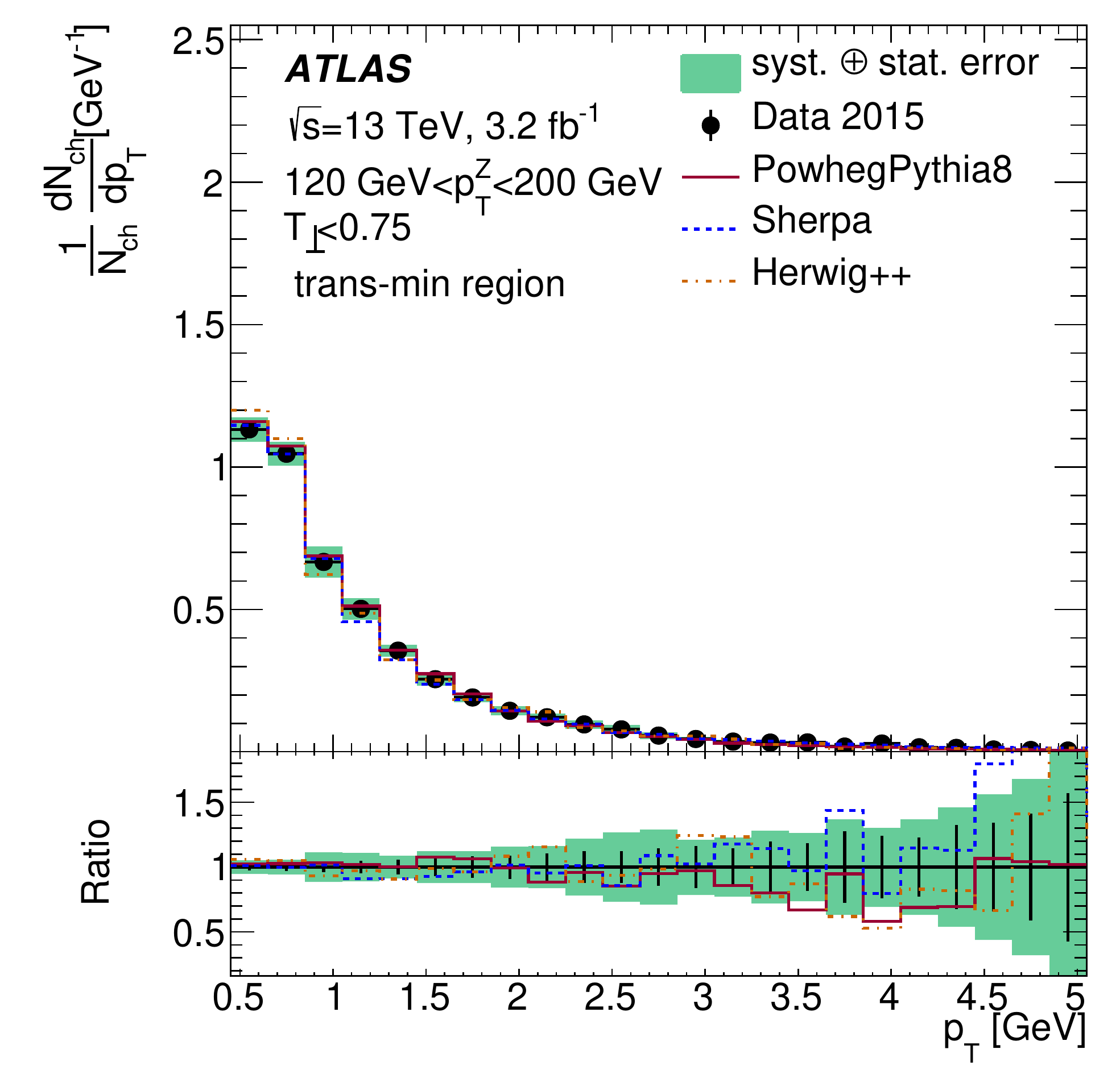}
\includegraphics[width=0.49\textwidth]{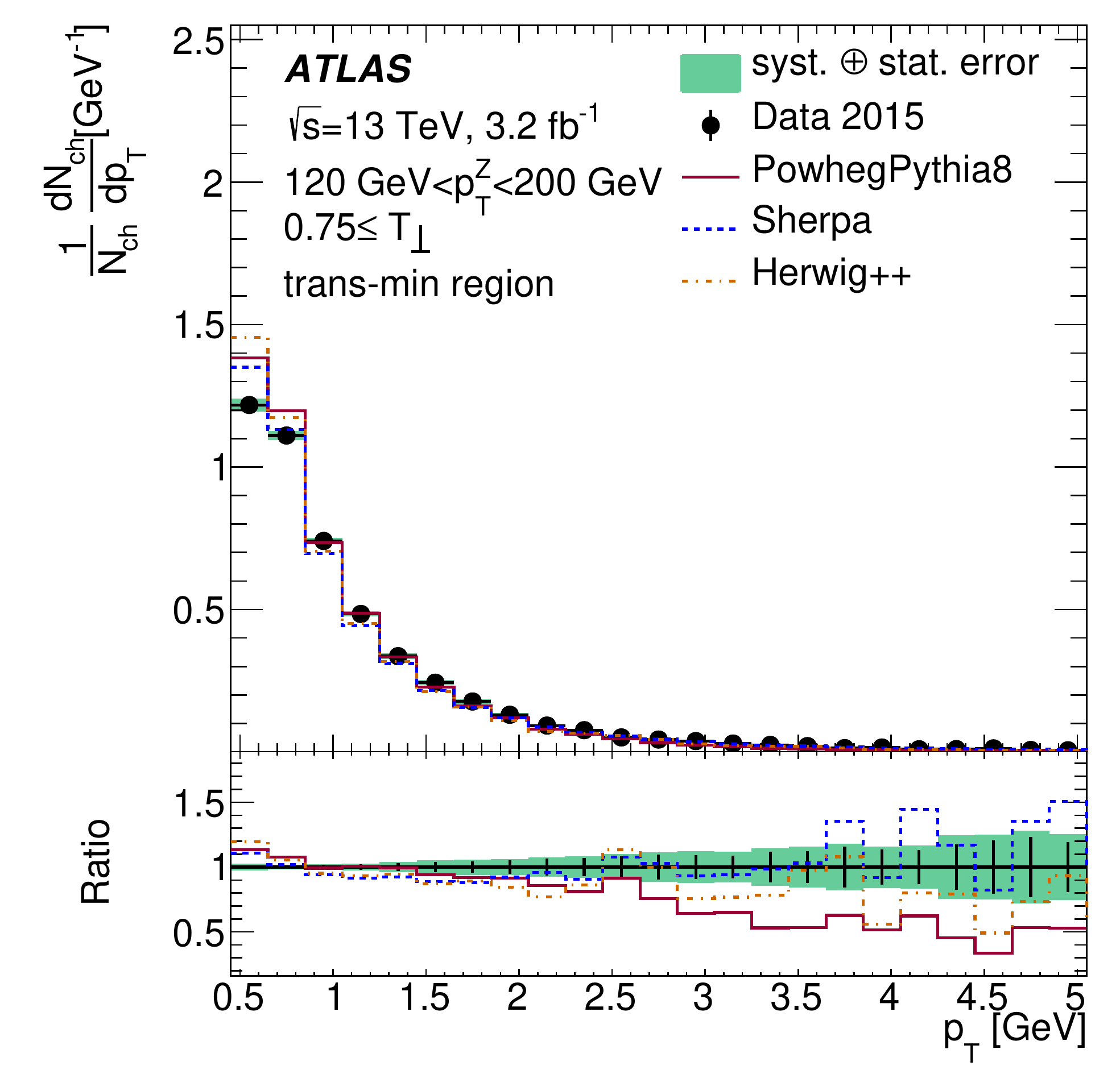}
\caption{Measured \particlept{} spectra in the trans-min region for $\thrust<0.75$ (left) and $0.75\leq\thrust$ (right) for events with 10 $<$ \Zpt{} $<$ 20~\GeV{} (upper row) and 120 $<$ \Zpt{} $<$ 200~\GeV{} (lower row).  Predictions of \Powheg{}+\Pythia{}, \Sherpa{}, and \HERWIGpp{} are compared with the data. The ratios shown are predictions over data.}
\label{fig:results-individual-thrust}
\end{figure}
 
\begin{figure}[ht]
\centering
\includegraphics[width=0.49\textwidth]{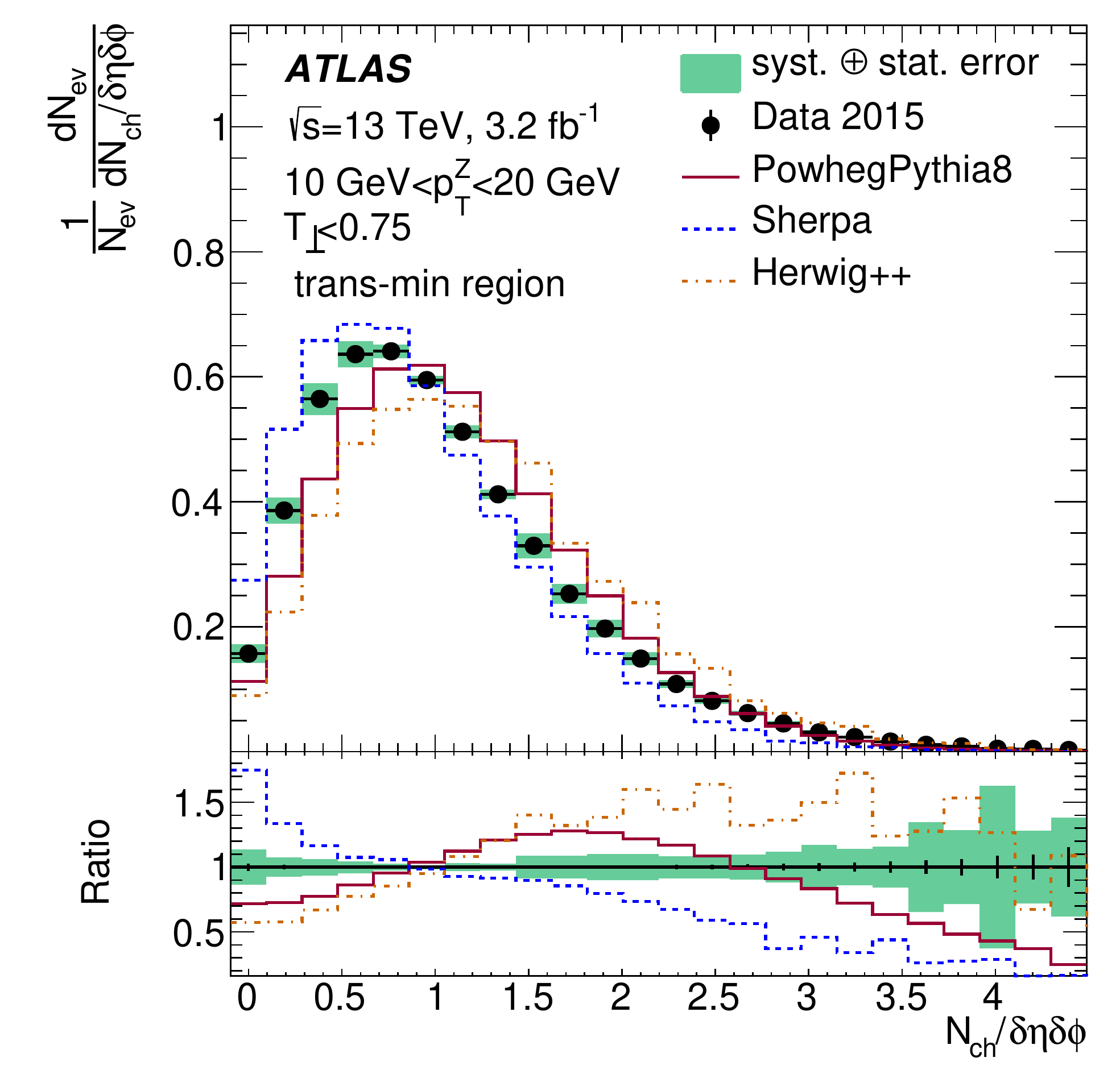}
\includegraphics[width=0.49\textwidth]{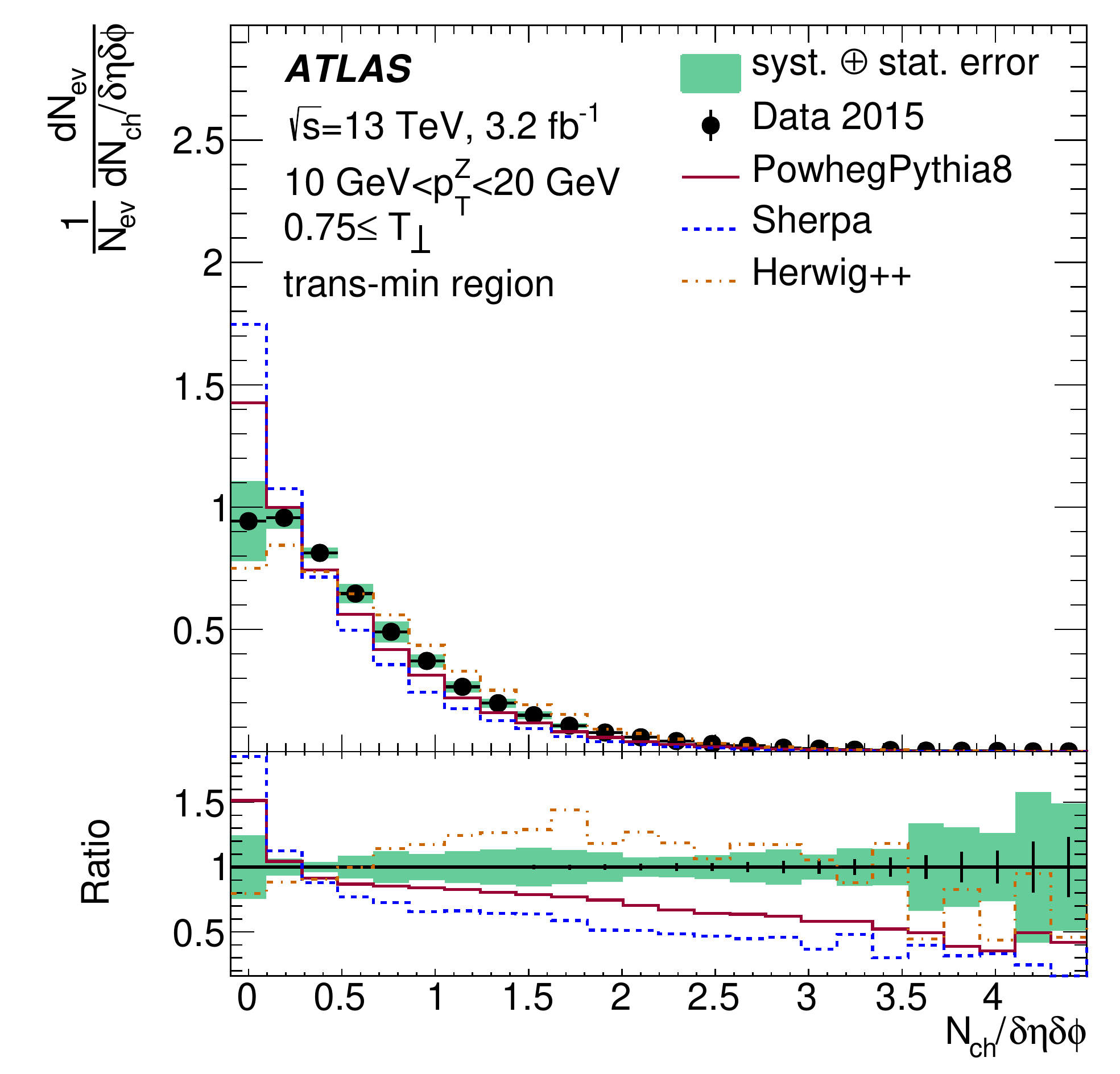}
\caption{Measured number of charged particles in the trans-min region for $\thrust<0.75$ (left) and $0.75\leq\thrust$ (right) for events with 10 $<$ \Zpt{} $<$ 20~\GeV{}.  Predictions of \Powheg{}+\Pythia{}, \Sherpa{}, and \HERWIGpp{} are compared with the data. The ratios shown are predictions over data.}
\label{fig:results-individual-thrust-nTracks}
\end{figure}
 
\subsection{Underlying-event activity as a function of \Zpt}
 
Figure~\ref{fig:pTZDependence} shows the mean number of charged particles and the mean of the scalar sum of the transverse momenta of those particles per unit $\eta$--$\phi$ space as a function of \Zpt{} in the transverse, trans-min, and trans-max regions inclusively in $\thrust$. The trans-min region is further separated by $\thrust$ in the right plots of Figure~\ref{fig:pTZDependence}.
In the trans-min region, the \UE-sensitive variables \nch{} and \sumpt{} rise slowly with increasing \Zboson{} boson transverse momentum. In contrast, the observables in the trans-max region have a strong dependence on \Zpt{}.
This is because it is heavily contaminated with the \Zboson{} boson hadronic recoil leaking into the transverse region.
The slope of the \UE\ activity in the trans-min region as a function of \Zpt{} for events of high \thrust{} is similar to the inclusive measurement.
The total amount of activity measured in the trans-min region for events with high \thrust{} is lower than the inclusive measurement due to the correlation of activity in the transverse region and \thrust.
Furthermore, the right-hand plots of Figure~\ref{fig:pTZDependence} demonstrate that the \UE\ activity is higher for events with lower \thrust{}, as expected \cite{Kar:2018uno}.
Lower values of \thrust{} also increase the dependence on \Zpt{} in the trans-min region.
 
\begin{figure}[ht]
\centering
\includegraphics[width=0.49\textwidth]{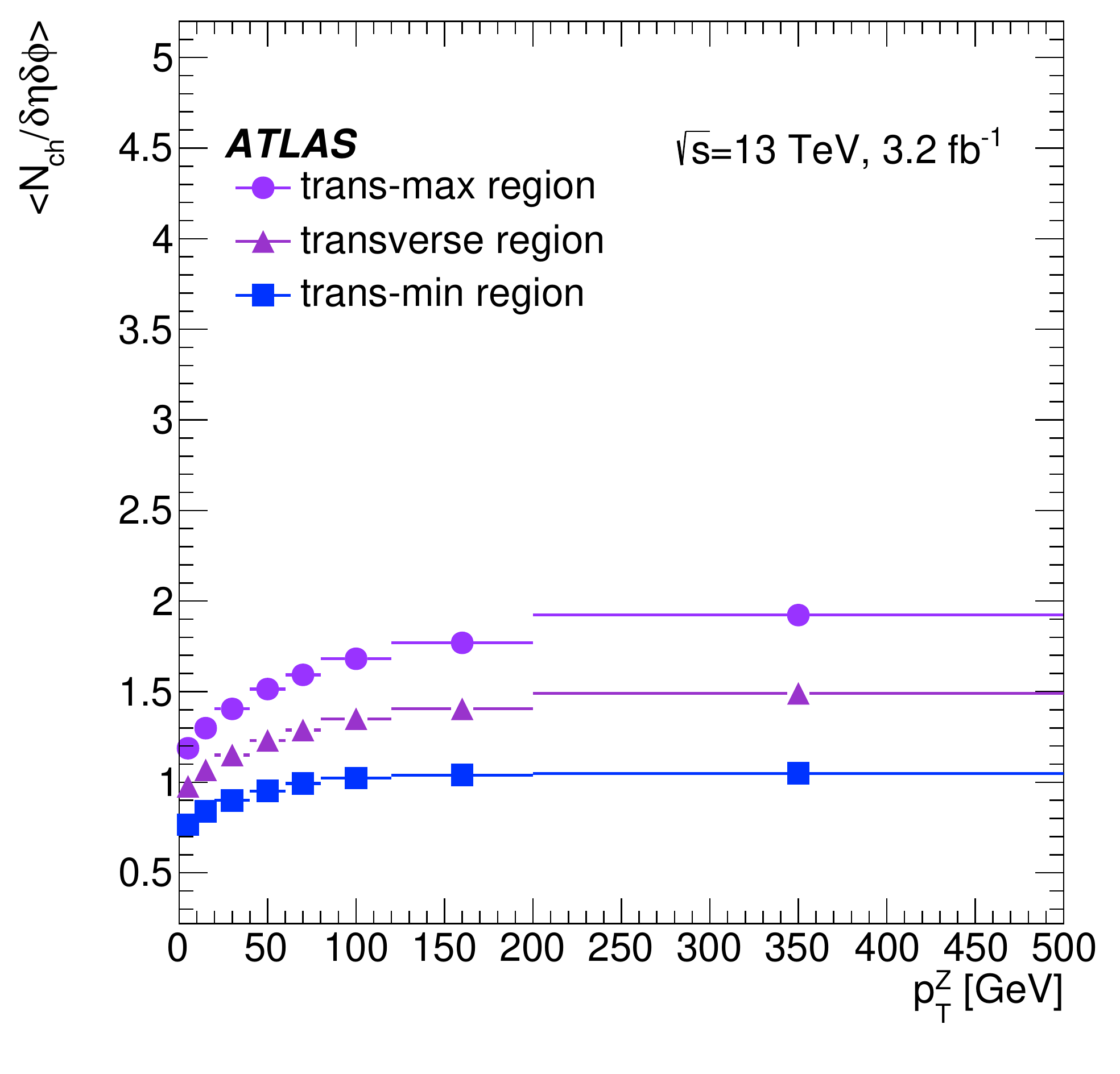}
\includegraphics[width=0.49\textwidth]{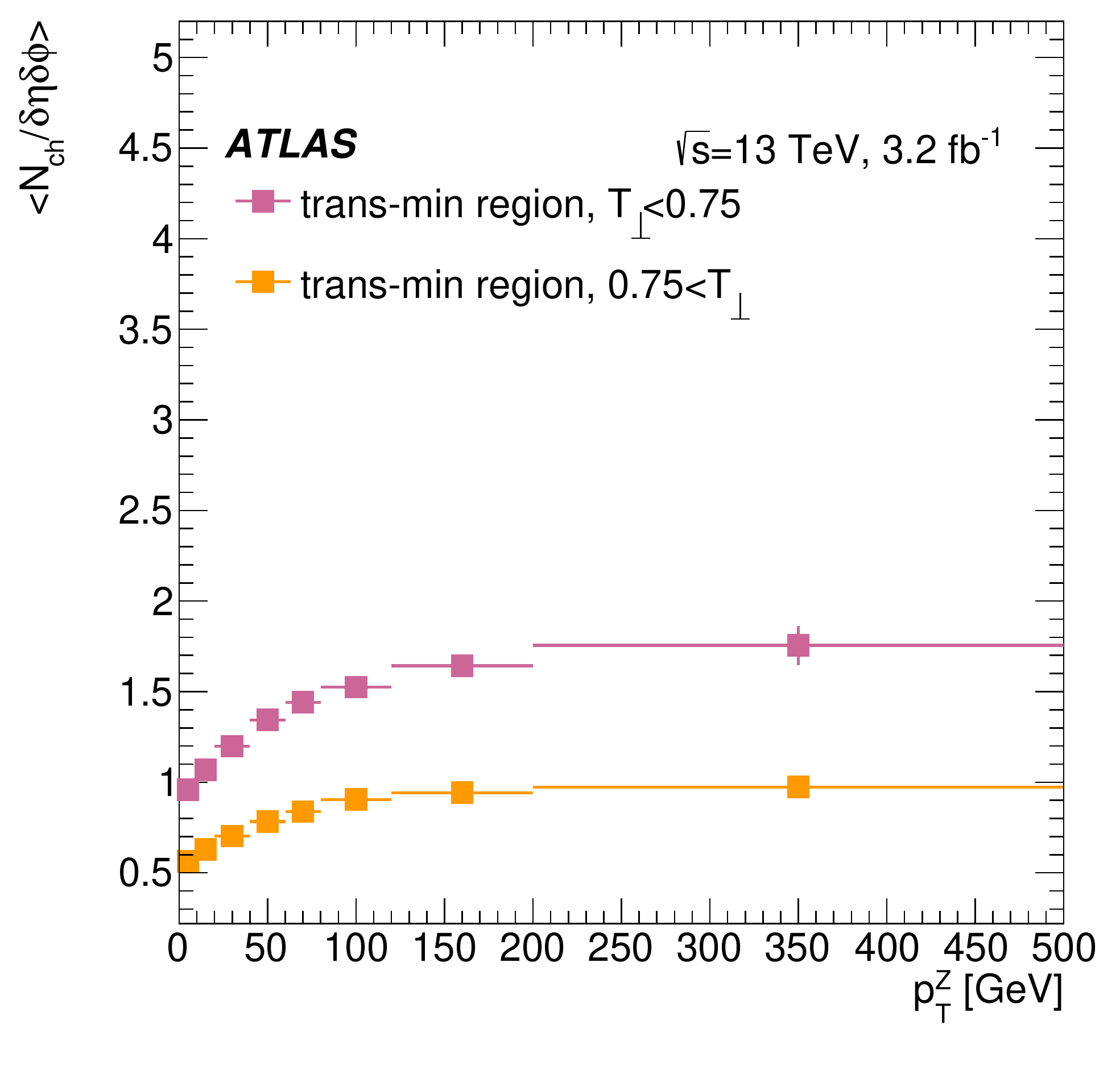}
\includegraphics[width=0.49\textwidth]{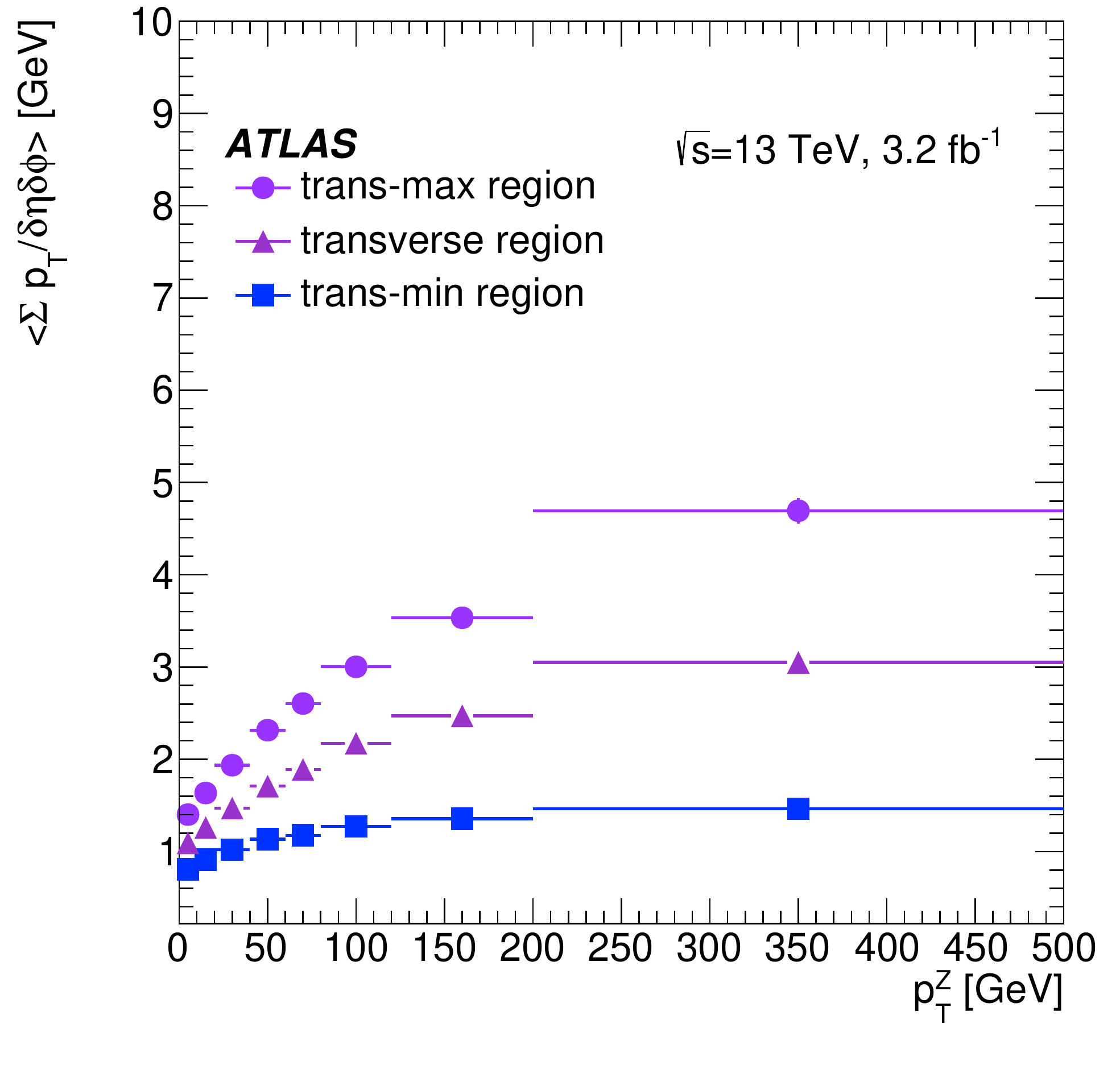}
\includegraphics[width=0.49\textwidth]{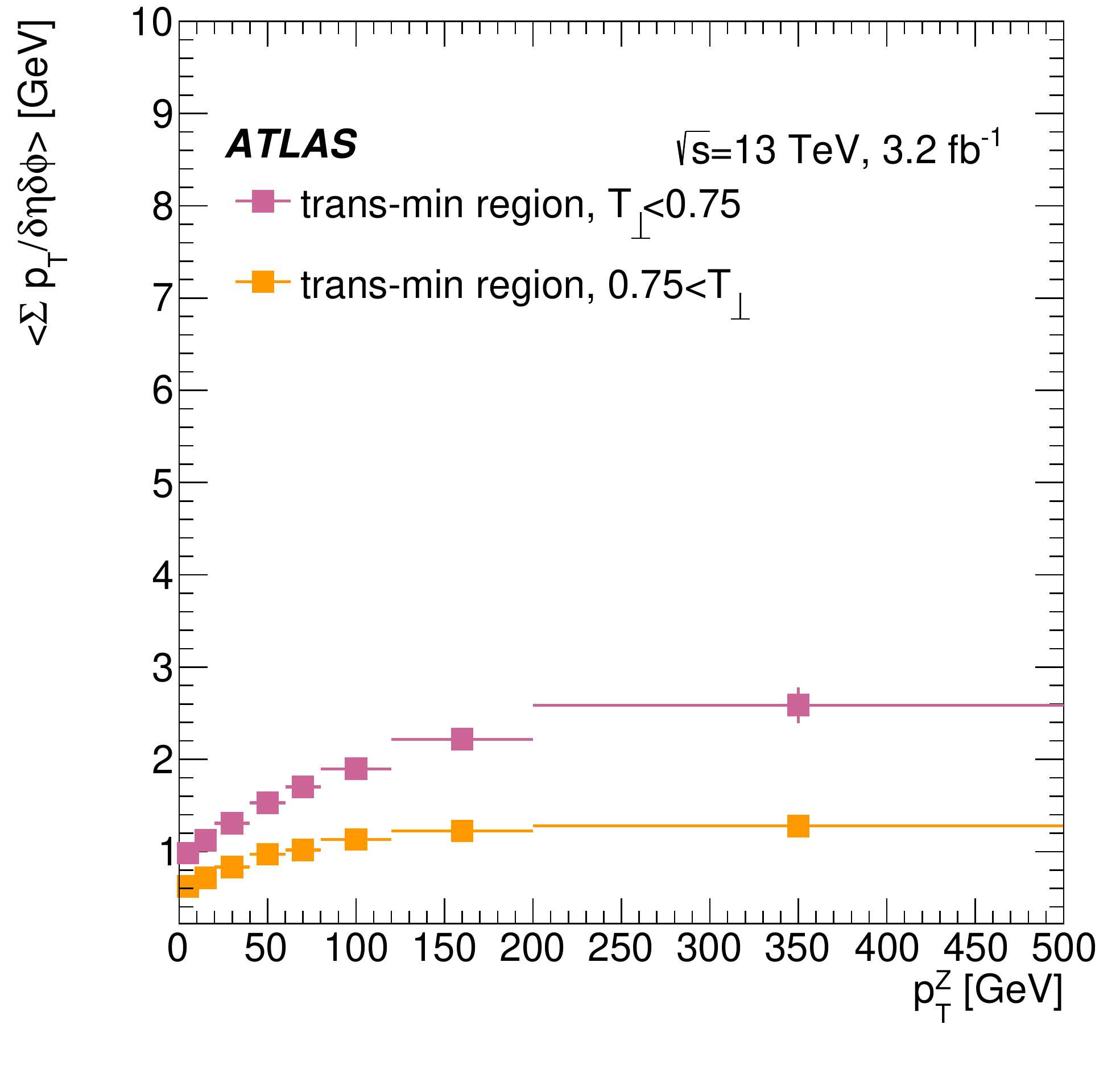}
\caption{The mean number of charged particles
(upper row) and the mean of the scalar sum of the transverse momentum of those particles (lower row) per unit $\eta$--$\phi$ space as a function of \Zpt{} in the full transverse region and for the trans-min and trans-max regions inclusively in $\thrust$ (left) and in the trans-min region separated in $\thrust$ (right).}
\label{fig:pTZDependence}
\end{figure}
 
The MC modelling of individual measurements in all 96
phase-space regions is further investigated by comparing the measured arithmetic means of the \nch{}, \sumpt{}, and \meanpt{} as functions of \Zpt{}.
Figures \ref{fig:results-pTZ-thrustIncl} and \ref{fig:results-meanpt} show comparisons with the predictions of \Powheg{}+\Pythia{}, \Sherpa{}, and \HERWIGpp{} for the trans-min and towards regions inclusively in $\thrust$.
The predictions fail to describe the data in either of the regimes.
For \Zpt\ $>$ 20~\GeV{}, \Herwigpp{} predicts a slower rise in \UE\ activity with rising \Zpt{} than in the measured distributions.
On the other hand, \Powheg{}+\Pythia{} and \Sherpa{} qualitatively describe the `turn-on' effect of the \UE\ activity, i.e.\ a steeper slope at low \Zpt{} which vanishes at higher values of \Zpt.
For \Powheg{}+\Pythia{}, the rise of the \UE\ activity is underestimated, and hence the discrepancy with data grows with \Zpt{} and stabilizes around \Zpt{} = 100~\GeV. Only in the toward region of the mean of the \meanpt\, is \Sherpa{} in good agreement with the data.
 
\begin{figure}[ht]
\centering
\includegraphics[width=0.49\textwidth]{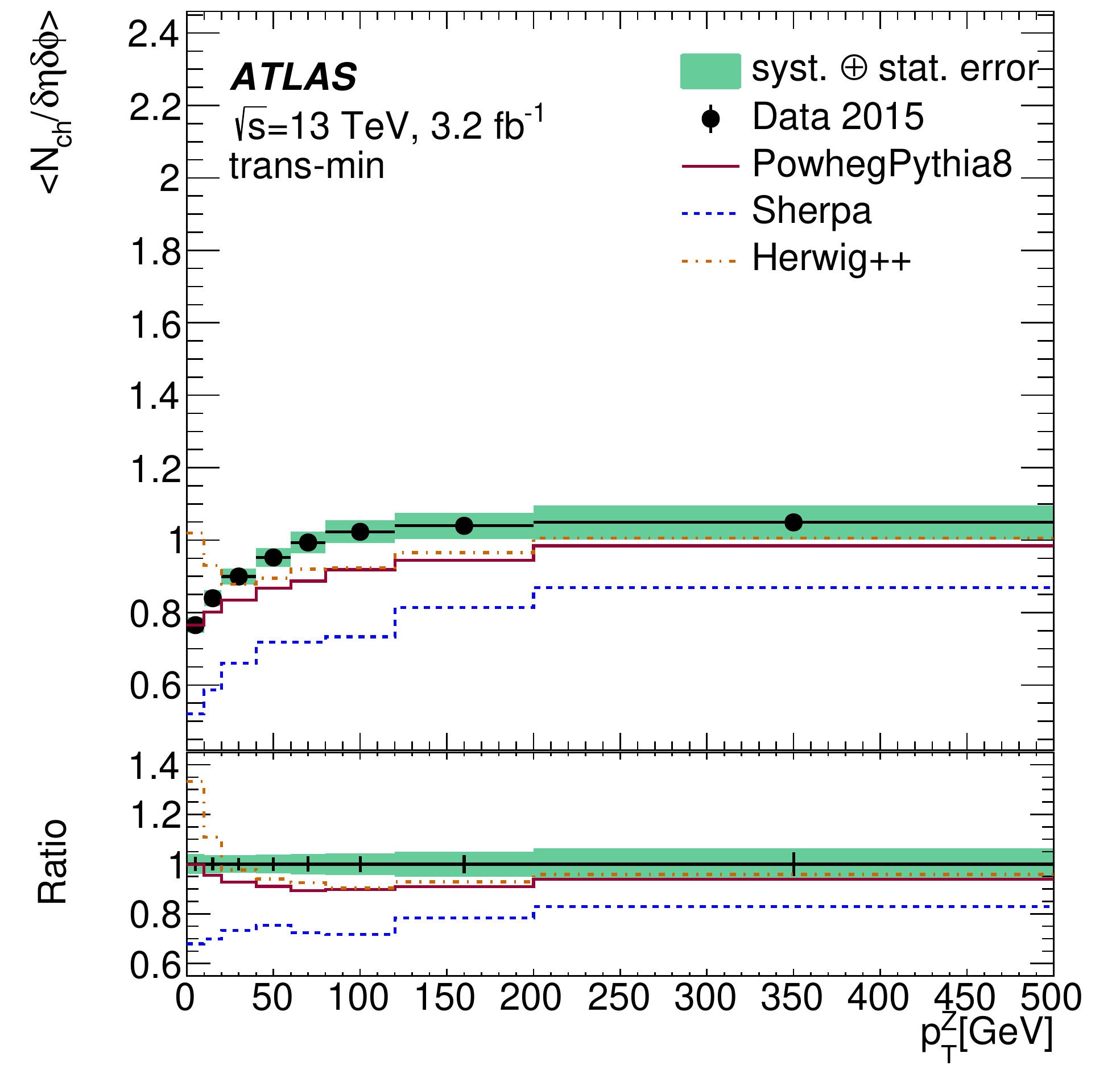}
\includegraphics[width=0.49\textwidth]{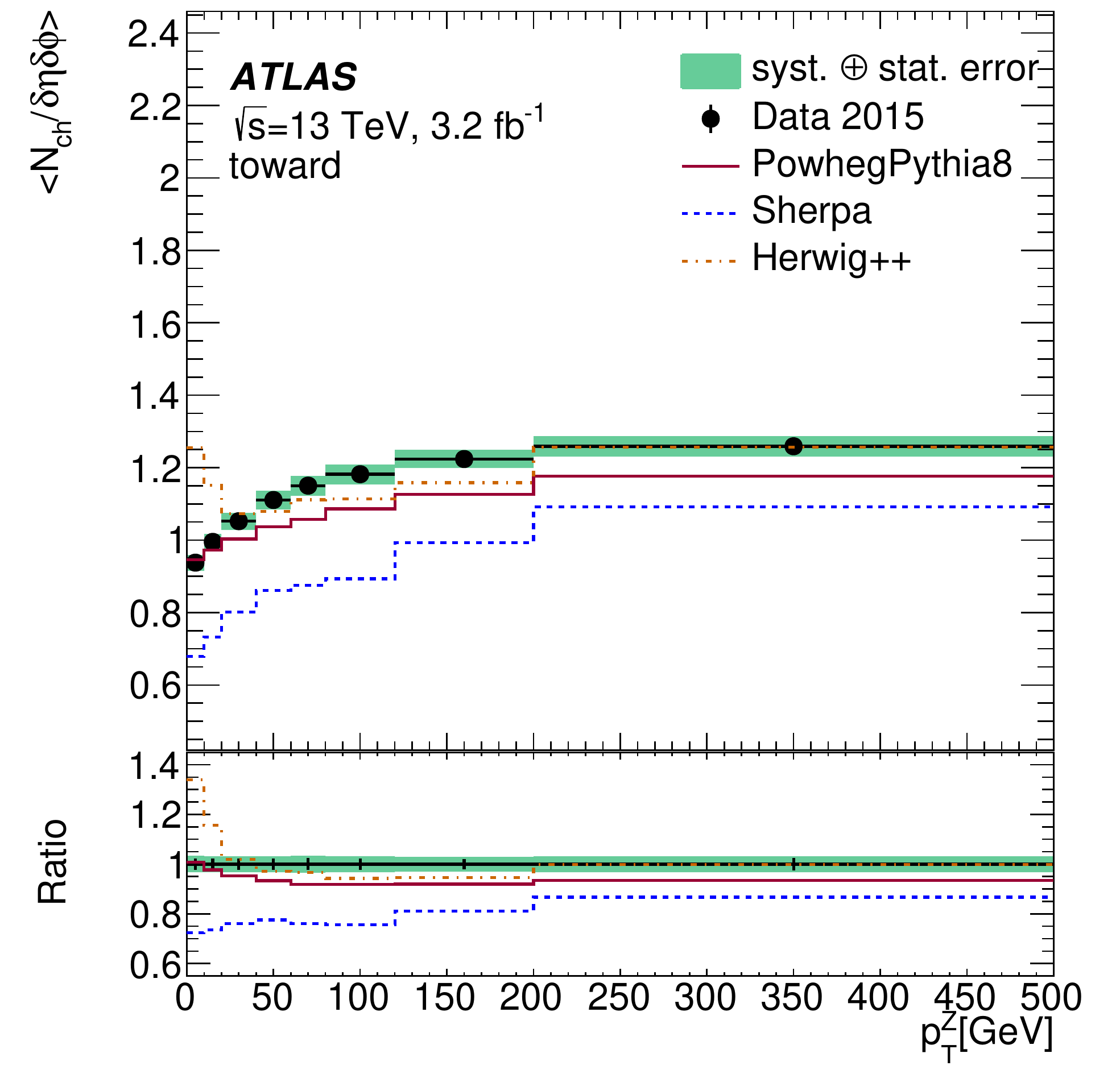}
\includegraphics[width=0.49\textwidth]{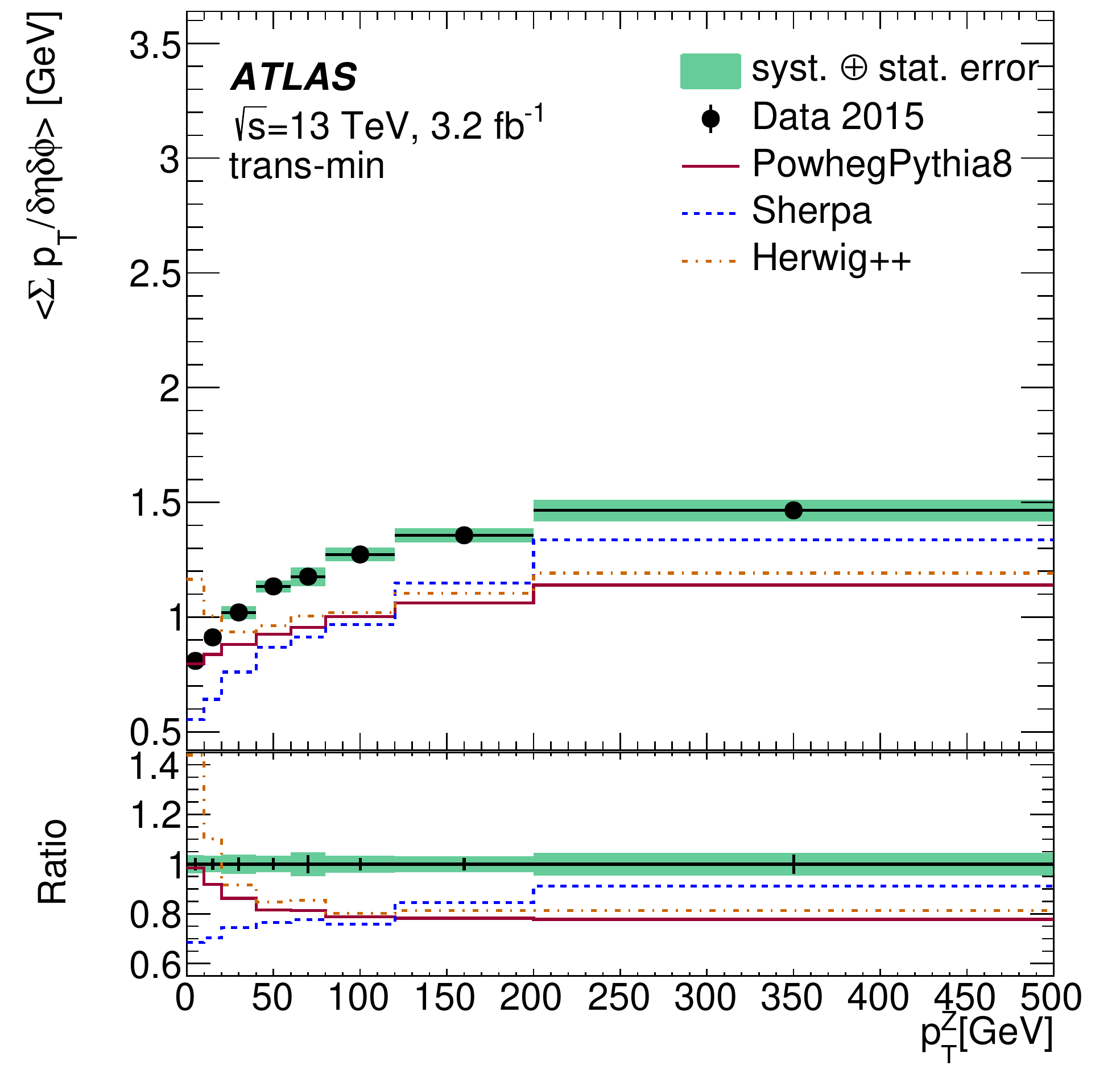}
\includegraphics[width=0.49\textwidth]{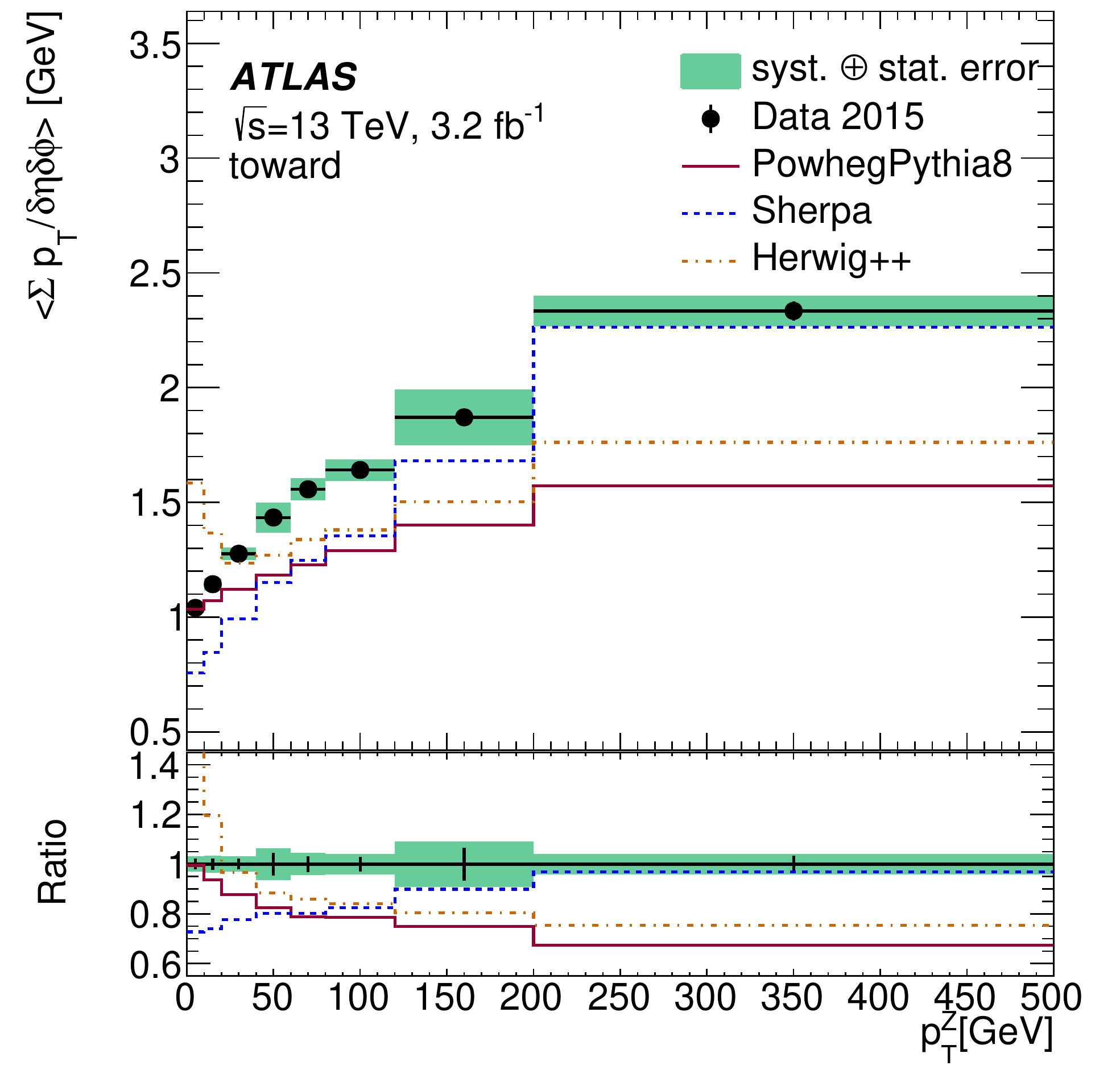}
\caption{Comparison of measured arithmetic means of the \nch{} (upper row) and \sumpt{} (lower row)
as functions of \Zpt{} for the trans-min (left) and towards (right) region inclusively in \thrust{}.  Predictions of \Powheg{}+\Pythia{}, \Sherpa{} and \HERWIGpp{} are compared with the data. The ratios shown are predictions over data.}
\label{fig:results-pTZ-thrustIncl}
\end{figure}
 
The \Zpt{} dependence for the two regions of $\thrust$ in the trans-min region is summarized in Figures~\ref{fig:results-pTZ-thrustRegions} and~\ref{fig:results-meanpt-thrust}. In the low $\thrust$ region, the prediction by \Sherpa{} improves, e.g.\ for \nch{} the discrepancy shrinks from about 30\% to roughly 10\%. Referring to the same observable, \Powheg{}+\Pythia{} is in agreement with data for \Zpt{} $>$ 80~\GeV{} in the low $\thrust$ regime within the uncertainties. For the selection on high $\thrust$ all generators underestimate the \UE\ activity.
\Sherpa{}~provides the best description of the data in $\langle \meanpt{}\rangle$.
Apart from the toward region, it tends to a constant underestimation but agrees with the overall shape.
The agreement of \Powheg{}+\Pythia{} with data is better for \thrust{} $<$ 0.75 than for the inclusive measurement.
The predictions of \Herwigpp{} in the trans-min region improve with higher values of \Zpt{} and also in events of lower \thrust{}. However, the discrepancy between \HERWIGpp{} and the data in the lowest bins remains regardless of the selected region.
 
\begin{figure}[ht]
\centering
\includegraphics[width=0.49\textwidth]{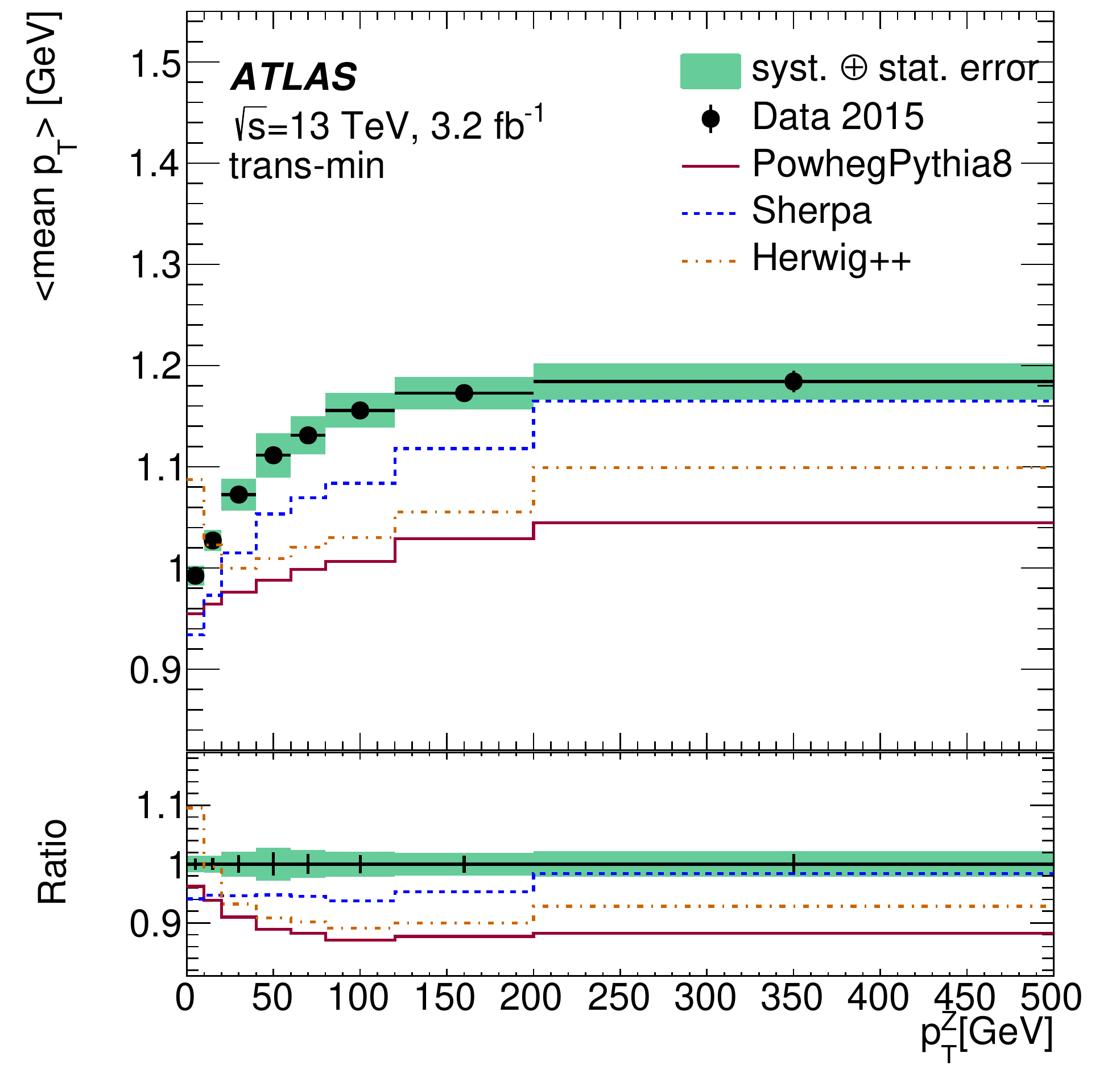}
\includegraphics[width=0.49\textwidth]{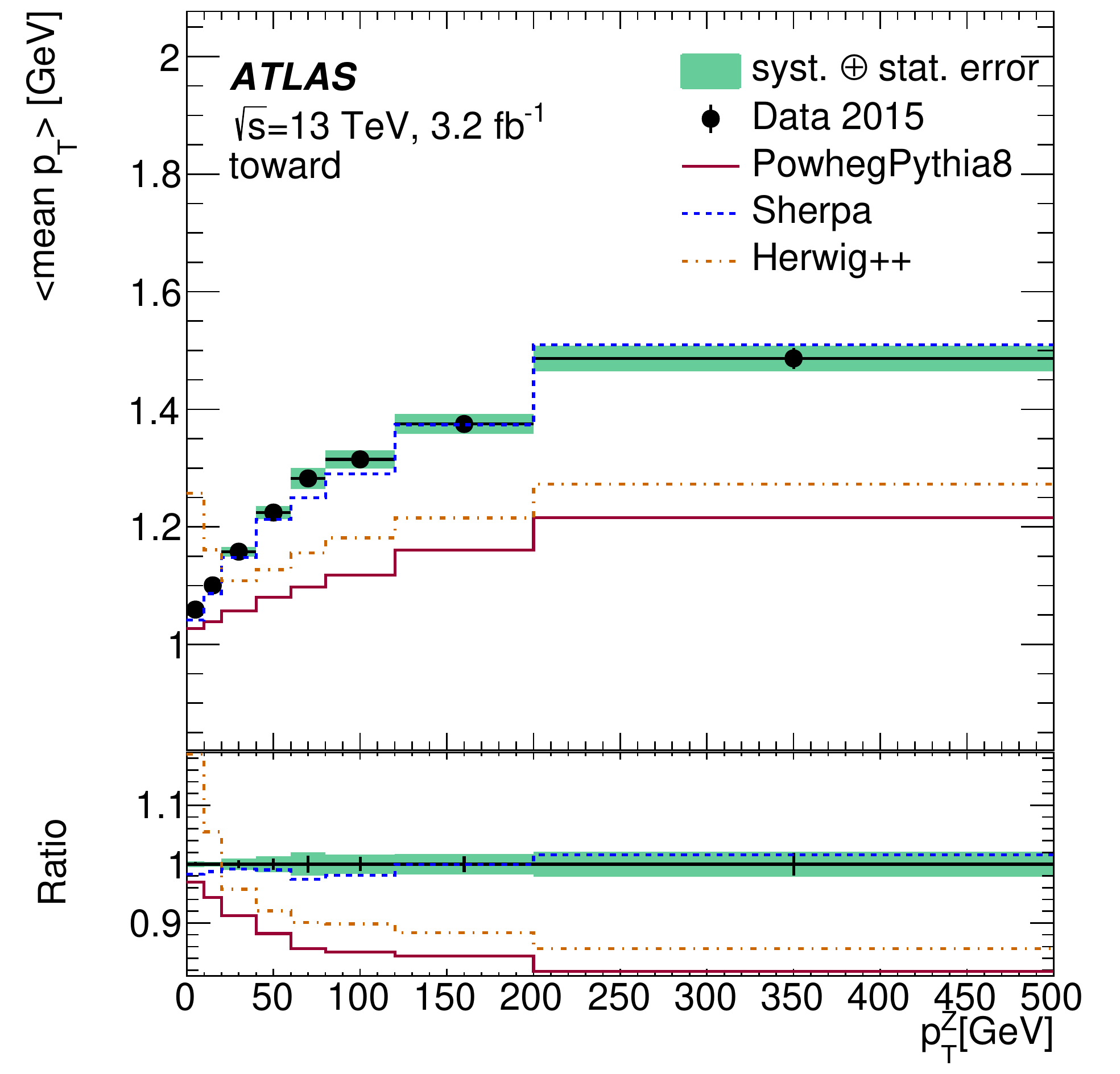}
\caption{Comparison of measured arithmetic means of \meanpt{} as functions of \Zpt{} for the trans-min (left) and towards (right) regions inclusively, and in regions of \thrust{}.  Predictions of \Powheg{}+\Pythia{}, \Sherpa{}, and \HERWIGpp{} are compared with the data. The ratios shown are predictions over data.}
\label{fig:results-meanpt}
\end{figure}
\begin{figure}[ht]
\centering
\includegraphics[width=0.49\textwidth]{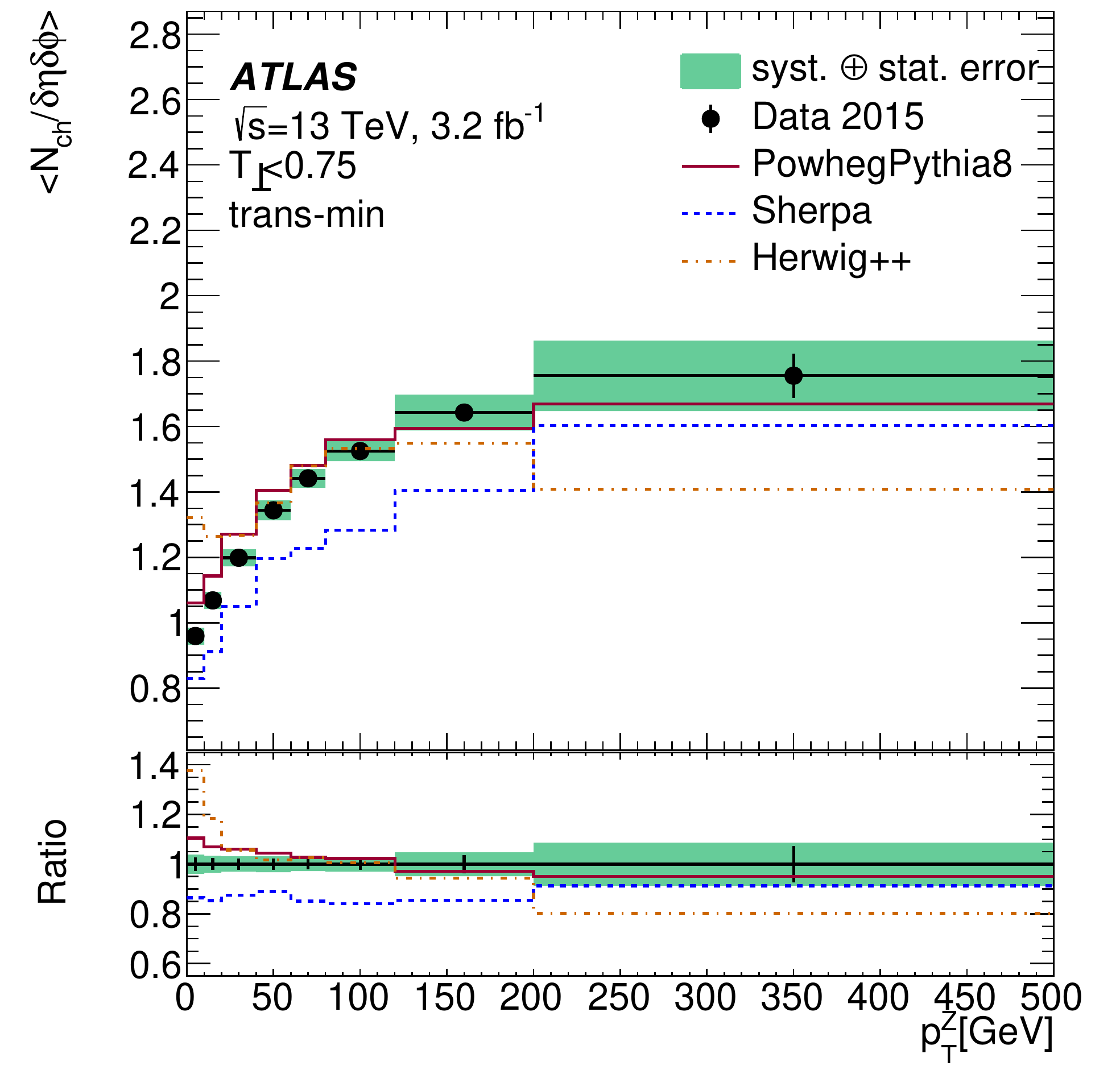}
\includegraphics[width=0.49\textwidth]{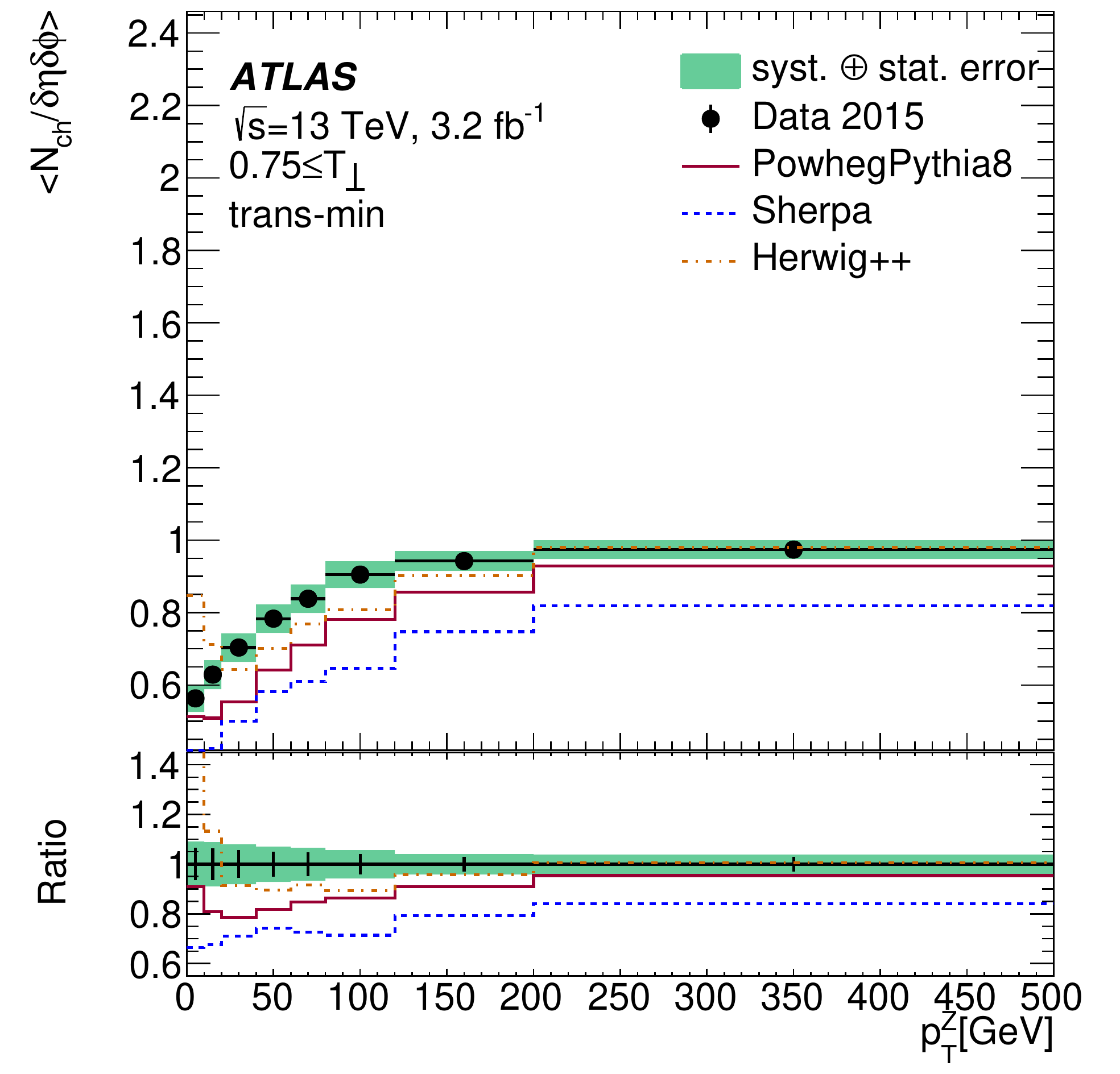}
\includegraphics[width=0.49\textwidth]{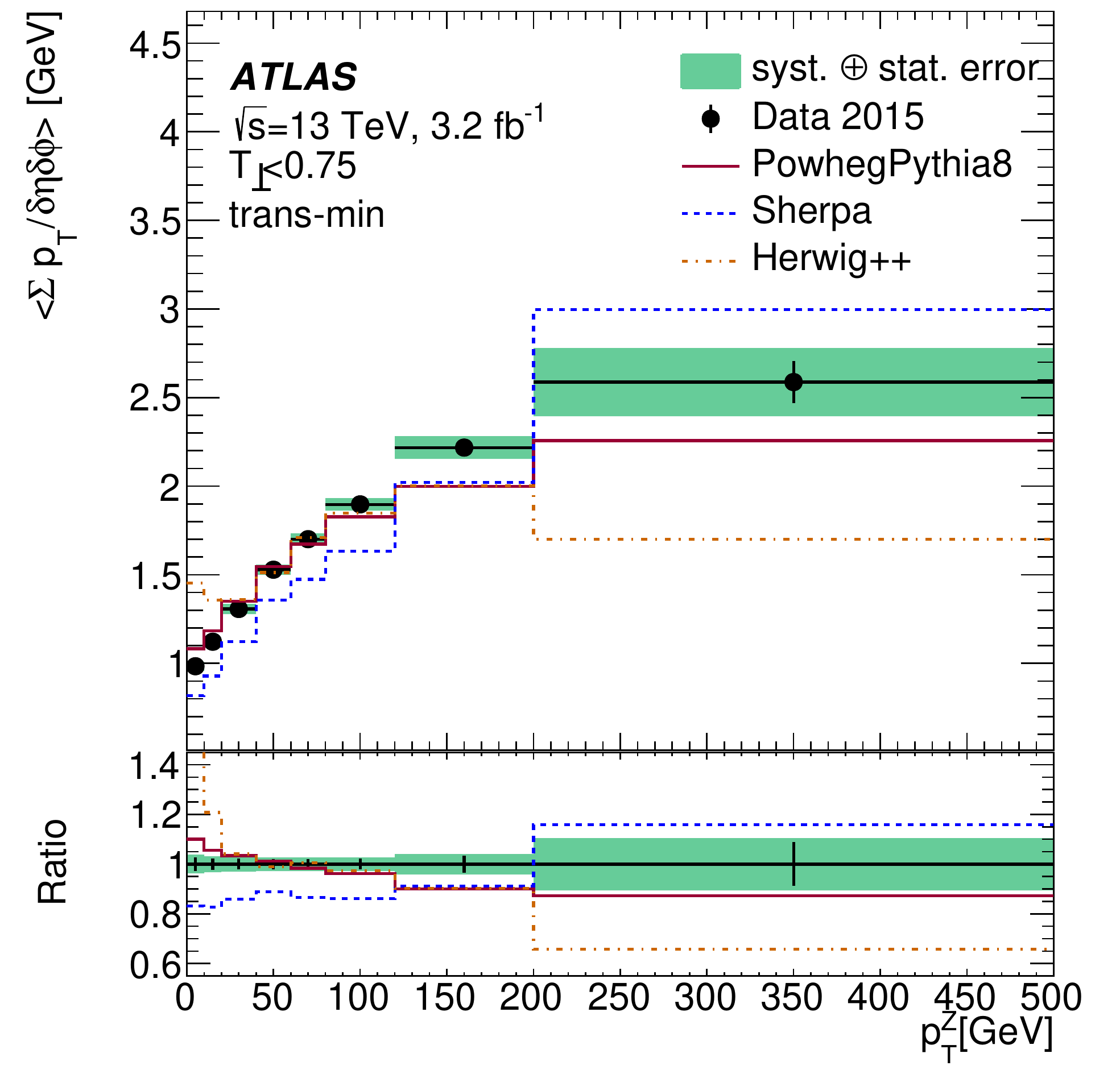}
\includegraphics[width=0.49\textwidth]{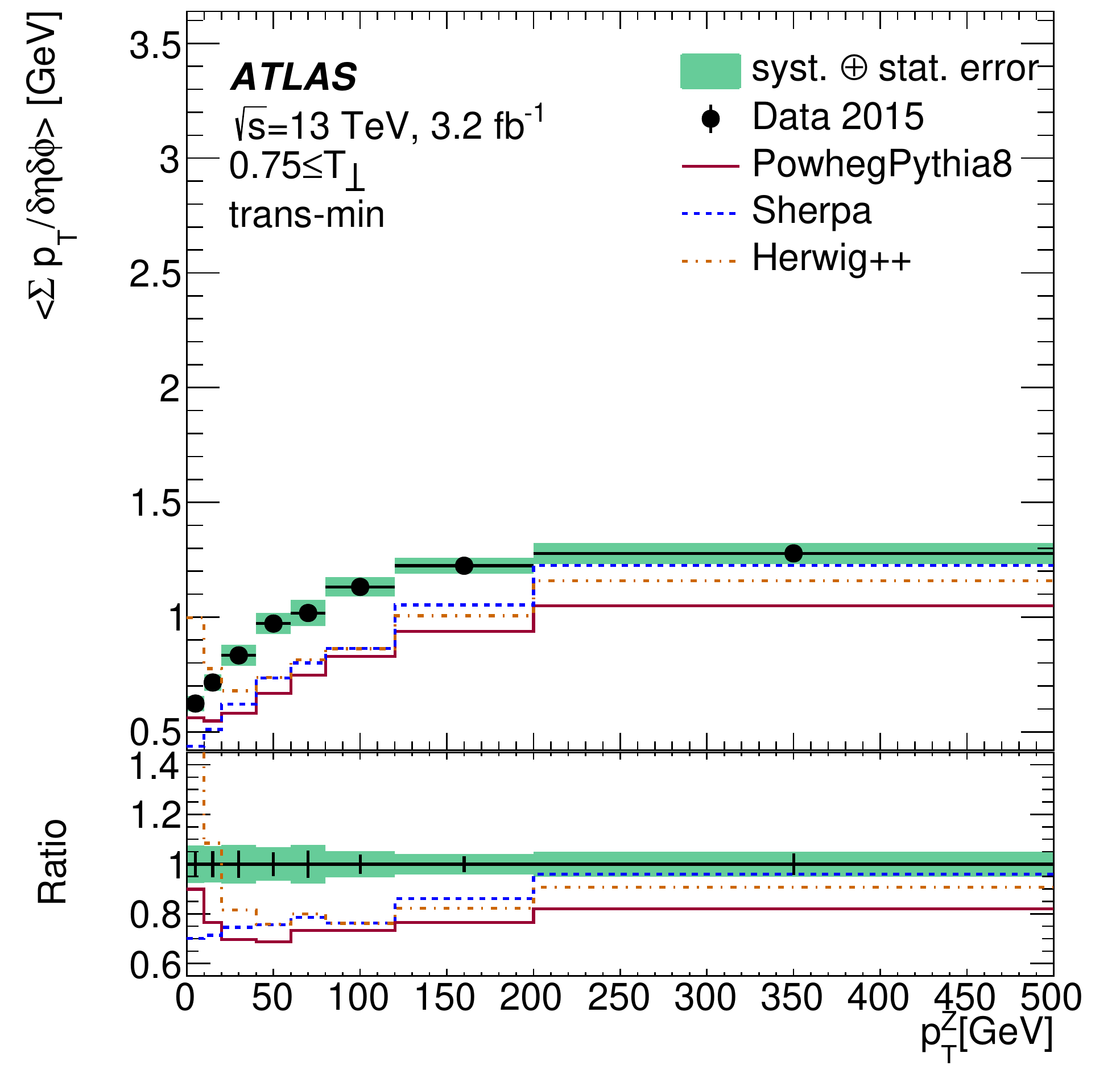}
\caption{Comparison of measured arithmetic means of the \nch{} (upper row) and \sumpt{} (lower row)
as functions of \Zpt{} for $\thrust{}<0.75$ (left) and $0.75\leq\thrust{}$ (right) for the trans-min region.  Predictions of \Powheg{}+\Pythia{}, \Sherpa{}, and \HERWIGpp{} are compared with the data. The ratios shown are predictions over data.}
\label{fig:results-pTZ-thrustRegions}
\end{figure}
 
\begin{figure}[ht]
\centering
\includegraphics[width=0.49\textwidth]{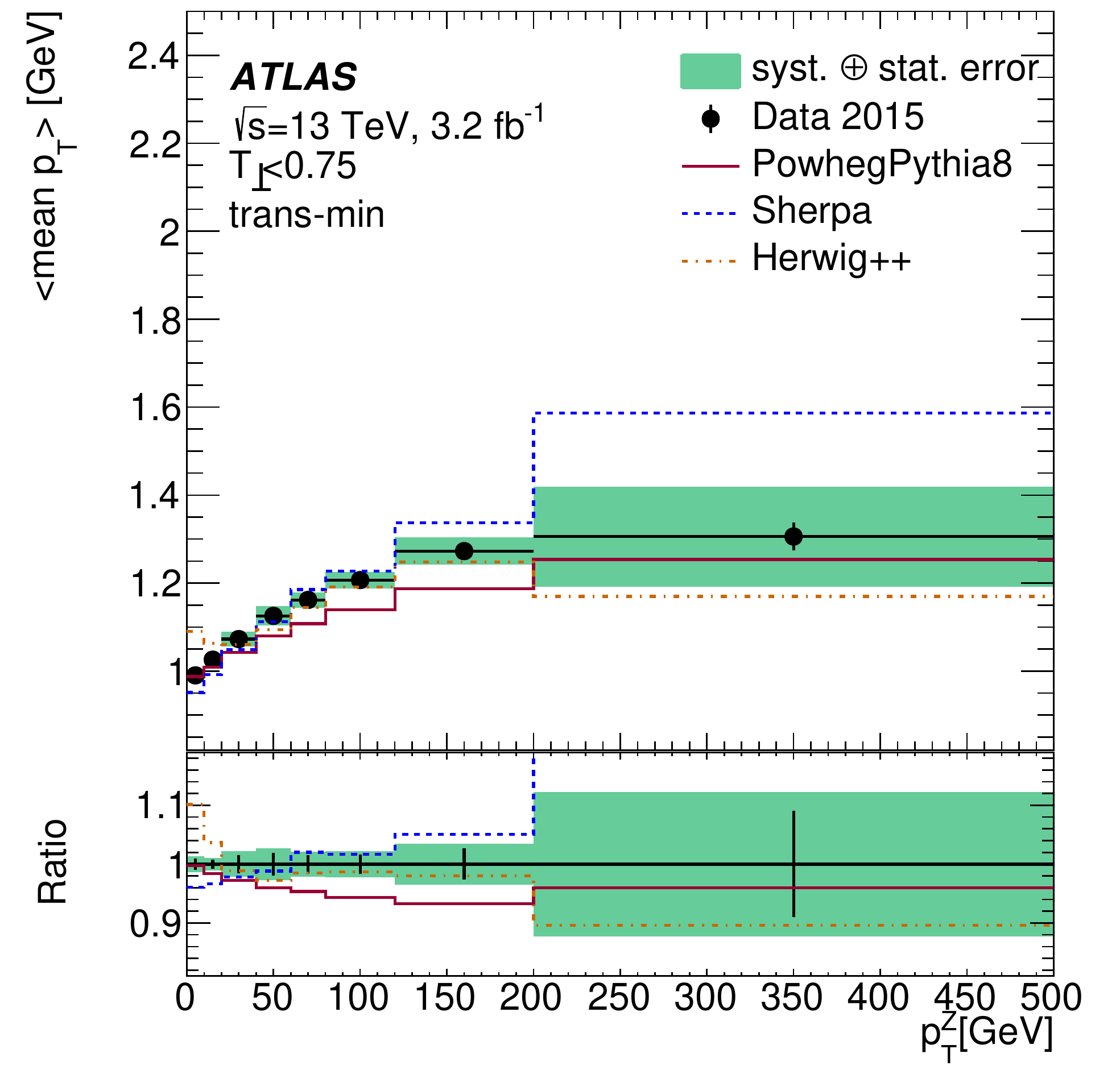}
\includegraphics[width=0.49\textwidth]{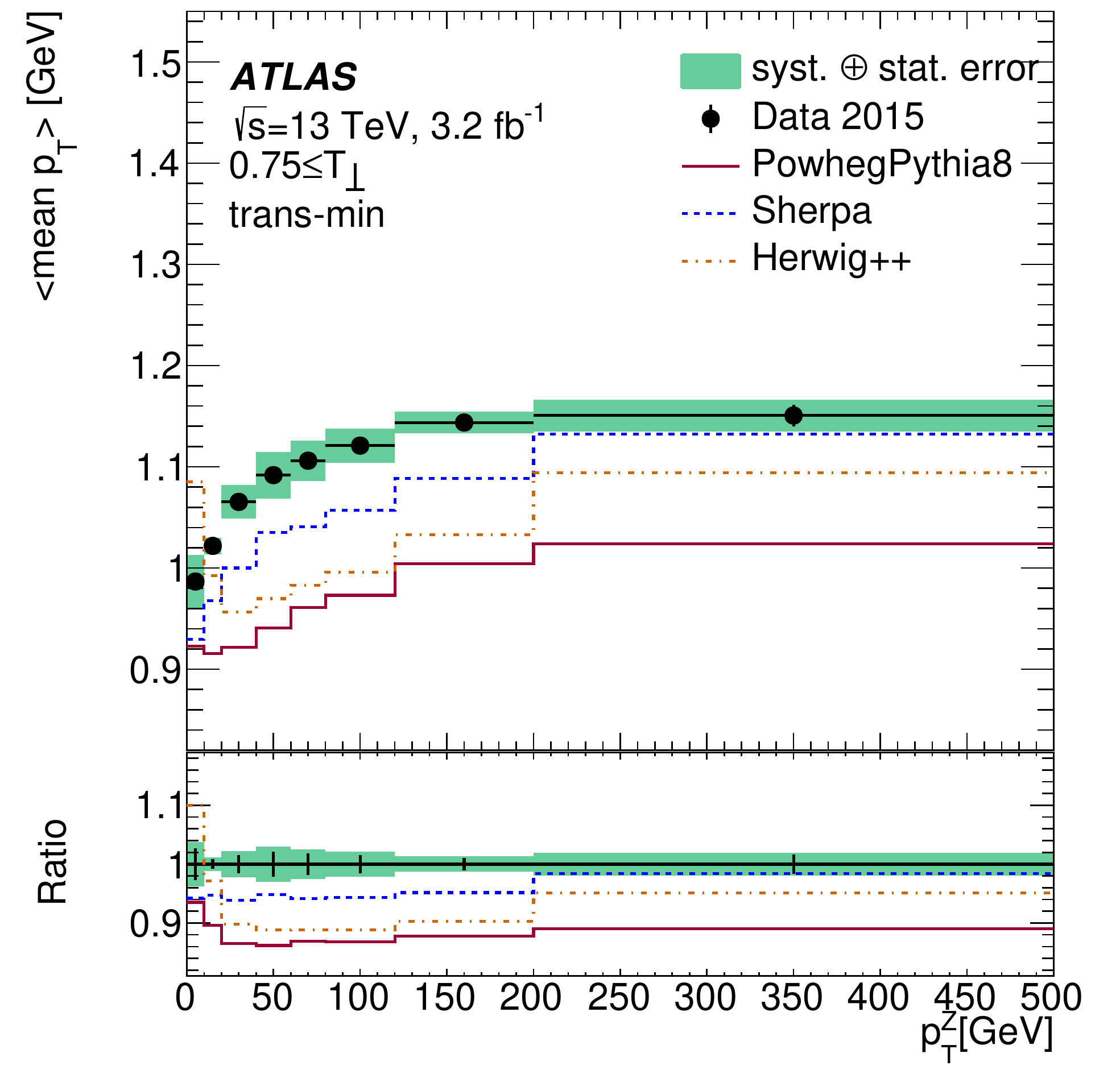}
\caption{Comparison of the measured arithmetic mean of \meanpt{} as a function of \Zpt{} for ranges of $\thrust$ in the trans-min region.  Predictions of \Powheg{}+\Pythia{}, \Sherpa{}, and \HERWIGpp{} are compared with the data. The ratios shown are predictions over data.}
\label{fig:results-meanpt-thrust}
\end{figure}
 
\subsection{Comparison with other centre-of-mass energies}
 
Figure \ref{fig:results-cdf7TeV} presents a comparison of the measured \Mnch{} and \Msumpt{} for different centre-of-mass energies.
The results for \rootseven{} are taken from the previous ATLAS measurement of the \UE\ activity in \Zboson{} boson events \cite{STDM-2011-42}. The event selection criteria are similar to the analysis presented in this paper, but the previous measurement also includes the $Z\rightarrow e^{+}e^{-}$ channel.
The CDF measurements at \rootonenine{}~\cite{Aaltonen:2010rm} are also included in the comparison. The CDF analyses used Drell--Yan lepton pairs in a smaller invariant mass window ($70<m_{\mu\mu}<110$~\GeV) in $p\bar{p}$ collisions.
The relative uncertainties of the two ATLAS measurements are of similar sizes, while the CDF measurements have large statistical fluctuations for $p_\mathrm{T}^{Z/\mu\mu}> 30\,\GeV$.
All three measurements show qualitatively the same behaviour, i.e.\ a growing \UE\ activity with higher values of \Zpt. 
With higher centre-of-mass energies, more energy is available for the processes forming the \UE\, e.g. MPI. Hence, the rise of the \UE\ activity as a function of \roots{} is expected.
 
\begin{figure}[ht]
\centering
\includegraphics[width=0.49\textwidth]{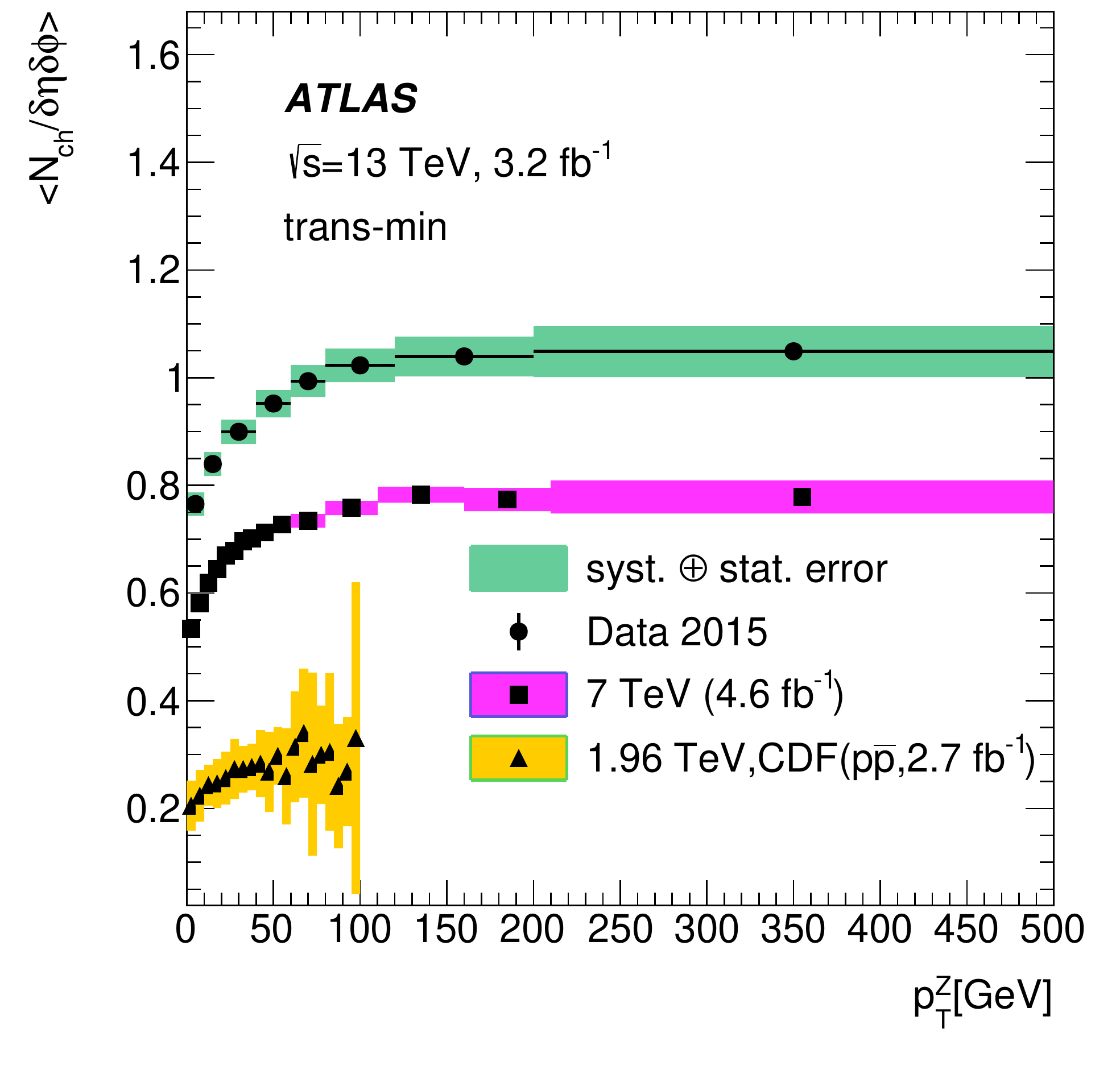}
\includegraphics[width=0.49\textwidth]{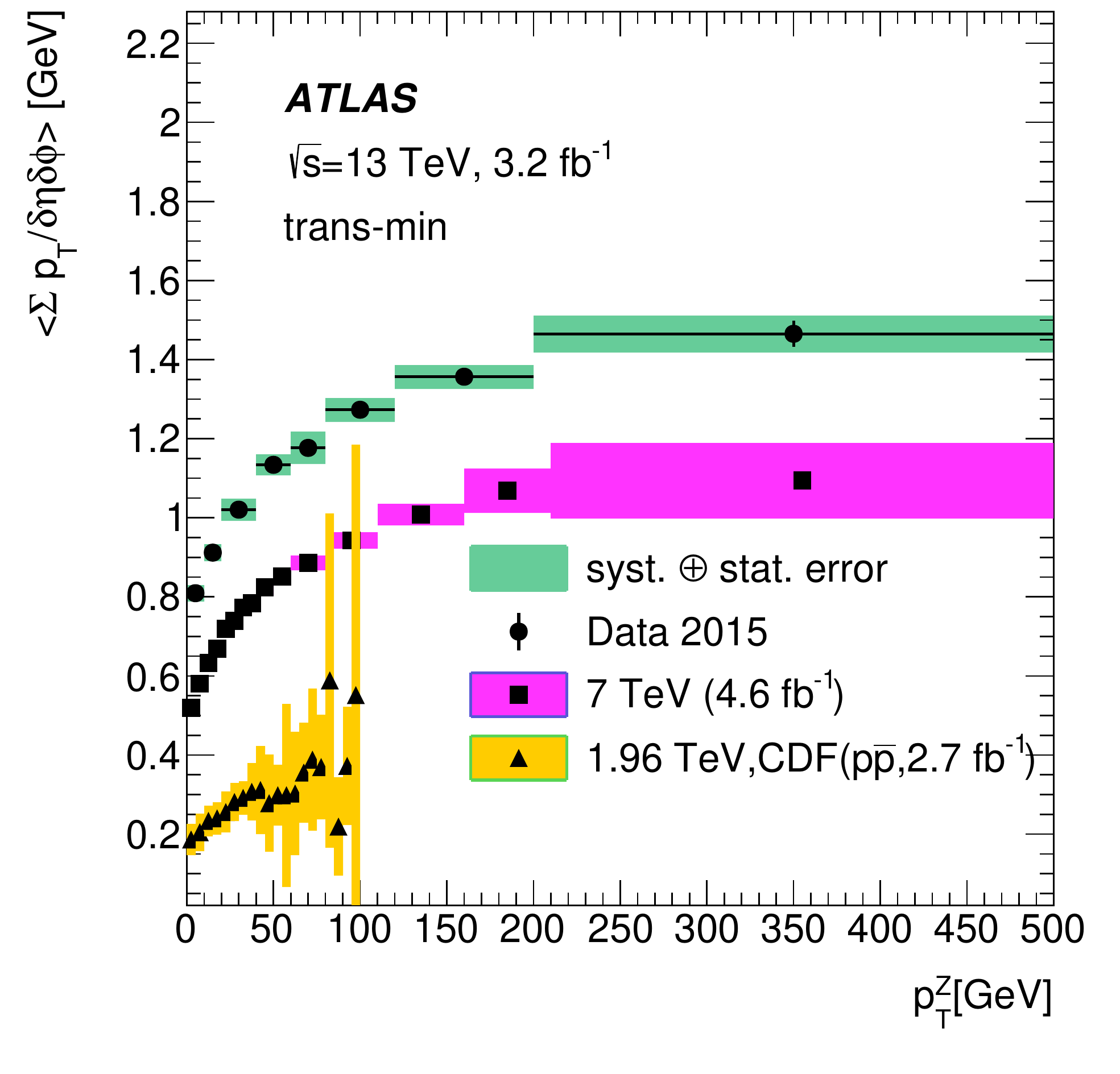}
\caption{The distributions of \Mnch{} and \Msumpt~measured at $\roots{}=\SI{13}{\tera\electronvolt}$ compared with the results of the previous ATLAS measurements at $\roots{}=\SI{7}{\tera\electronvolt}$~\cite{STDM-2011-42} and the CDF measurements at \roots{}=\SI{1.96}{\tera\electronvolt}~\cite{Aaltonen:2010rm}. The error bars correspond to the full uncertainties of the corresponding measurement.}
\label{fig:results-cdf7TeV}
\end{figure}
\FloatBarrier
 
\section{Discussion and conclusion}
 
Measurements of four observables sensitive to the activity of the \UE\ in \Zmm{} events are presented using 3.2~\ifb{} of \sqs\ = 13~\TeV{} $pp$ collision data collected with the ATLAS detector at the LHC in 2015.
Those observables are the \pT{} of charged particles,
the number of charged particles per event (\nch{}), the sum of charged-particle \pT{} per event (\sumpt{}), and the mean of charged-particle \pT{} per event (\meanpt{}).
They are measured in intervals of the \Zboson{} boson \pt{}
and in different azimuthal regions of the detector relative to the \Zboson{} boson direction.
The arithmetic means of the distributions are plotted as functions of the \Zboson{} boson \pt{},
inclusively of and in regions of transverse thrust.
 
The predictions from three Monte Carlo generators (\Powheg{}+\Pythiaeight{}, \Sherpa{} and~\HERWIGpp{}) are compared with the data.
In general, all tested generators and tunes show significant deviations from the data distributions regardless of the observable.
The arithmetic means of the observables deduced from the predictions of \Powheg{}+\Pythiaeight{} and \Sherpa{} match the main features of the \UE\ activity in the fiducial region.
The turn-on effect, i.e.\ the rising activity as a function of the hard-scatter scale (here \Zpt), is visible as is a saturation of this effect for higher values of \Zpt.
In contrast to the other generators, \HERWIGpp{} fails to reproduce the turn-on effect at low \Zpt\ as it predicts that the \UE\ activity decreases as a function of \Zpt\ when considered only in the $\Zpt < 20\,\GeV{}$ region.
Otherwise, all generators underestimate the activity of the \UE\ when quantified as the arithmetic mean of the observables for inclusive \thrust{}.
The generators predict the mean values better in comparison with the data when focusing on the MPI-sensitive regions. \Powheg{}+\Pythiaeight{} is in agreement with data within the uncertainties for \Mnch~and \Msumpt,
indicating an adequate handling of the MPI activity. However, since the predictive power shrinks for the region with $\thrust>0.75$ in comparison with the inclusive measurement, the simulation of contributions other than MPI to the \UE\ activity needs to be improved.
Reference~\cite{Kar:2018uno} points out that the region with $\thrust>0.75$ is dominated by extra jet activity, giving a first indication for a possible improvement of the MC generator prediction.
This conclusion is valid when focusing on \Powheg{}+\Pythiaeight{} for different regions of \thrust\ for individual bins of \Zpt{}.
 
In comparison with the measurements at $\sqrt{s}$ = 7~\TeV~\cite{STDM-2011-42}, the performance of \HERWIGpp{} is consistent for \Zpt\ $>$ 20~\GeV. Both measurements use the energy-extrapolation tunes \cite{Gieseke:2012ft} provided by the \Herwigpp{} authors, i.e.\ \textsc{UE-EE-3} for \rootseven{} and in the analysis presented here \textsc{UE-EE-5}.
The latter tune was additionally validated against Tevatron and LHC measurements at $\sqrt{s}=\SI{900}{\giga\electronvolt}$ and \rootseven{} \cite{Seymour:2013qka}.
The prediction of \Herwigpp{} is slightly better for the distributions of \Mnch{} and \Msumpt{} at higher values of \Zpt. In the previous measurements, the divergence increased with \Zpt, which might be related to improper modelling of the impact parameter.
Apart from overestimating the mean activity, \Herwigpp{} improved relative to the \rootseven{} measurements in the description of the shape of \dsumpt{}, \dmeanpt{}, and \dnch{} in the presented \Zpt{}-bins. Qualitatively it performs better than the other generators.
 
\Powheg{}+\Pythiaeight{} performs as well at \rootthirteen{} as it does at \rootseven{}, but is tuned with \textsc{AU2} (only the MPI part was tuned by ATLAS using \rootseven{} \UE\ data) in the previous measurements. Nevertheless, this indicates that the MPI energy extrapolation of \Pythiaeight{} works well, which is in agreement with the better description for distributions at low \thrust.
 
In contrast, while at \rootseven{} \Sherpa{} version 1.4.0 with the \textsc{CT10} PDF set consistently overestimates the \UE{} activity metrics \Mnch{} and \Msumpt{} by 5\% to 15\%, the present analysis and \Sherpa{} version reveal a continuous underestimation. At \rootthirteen{}, the discrepancy relative to the data decreases with higher values of \Zpt.
 
\section*{Acknowledgements}
 
We thank CERN for the very successful operation of the LHC, as well as the
support staff from our institutions without whom ATLAS could not be
operated efficiently.
 
We acknowledge the support of ANPCyT, Argentina; YerPhI, Armenia; ARC, Australia; BMWFW and FWF, Austria; ANAS, Azerbaijan; SSTC, Belarus; CNPq and FAPESP, Brazil; NSERC, NRC and CFI, Canada; CERN; CONICYT, Chile; CAS, MOST and NSFC, China; COLCIENCIAS, Colombia; MSMT CR, MPO CR and VSC CR, Czech Republic; DNRF and DNSRC, Denmark; IN2P3-CNRS, CEA-DRF/IRFU, France; SRNSFG, Georgia; BMBF, HGF, and MPG, Germany; GSRT, Greece; RGC, Hong Kong SAR, China; ISF and Benoziyo Center, Israel; INFN, Italy; MEXT and JSPS, Japan; CNRST, Morocco; NWO, Netherlands; RCN, Norway; MNiSW and NCN, Poland; FCT, Portugal; MNE/IFA, Romania; MES of Russia and NRC KI, Russian Federation; JINR; MESTD, Serbia; MSSR, Slovakia; ARRS and MIZ\v{S}, Slovenia; DST/NRF, South Africa; MINECO, Spain; SRC and Wallenberg Foundation, Sweden; SERI, SNSF and Cantons of Bern and Geneva, Switzerland; MOST, Taiwan; TAEK, Turkey; STFC, United Kingdom; DOE and NSF, United States of America. In addition, individual groups and members have received support from BCKDF, CANARIE, CRC and Compute Canada, Canada; COST, ERC, ERDF, Horizon 2020, and Marie Sk{\l}odowska-Curie Actions, European Union; Investissements d' Avenir Labex and Idex, ANR, France; DFG and AvH Foundation, Germany; Herakleitos, Thales and Aristeia programmes co-financed by EU-ESF and the Greek NSRF, Greece; BSF-NSF and GIF, Israel; CERCA Programme Generalitat de Catalunya, Spain; The Royal Society and Leverhulme Trust, United Kingdom.
 
The crucial computing support from all WLCG partners is acknowledged gratefully, in particular from CERN, the ATLAS Tier-1 facilities at TRIUMF (Canada), NDGF (Denmark, Norway, Sweden), CC-IN2P3 (France), KIT/GridKA (Germany), INFN-CNAF (Italy), NL-T1 (Netherlands), PIC (Spain), ASGC (Taiwan), RAL (UK) and BNL (USA), the Tier-2 facilities worldwide and large non-WLCG resource providers. Major contributors of computing resources are listed in Ref.~\cite{ATL-GEN-PUB-2016-002}.

\printbibliography
 
\clearpage
 
\appendix

\clearpage
\input{atlas_authlist}\end{document}

%% file: atlas_authlist.tex
 
\begin{flushleft}
{\Large The ATLAS Collaboration}

\bigskip

G.~Aad$^\textrm{\scriptsize 101}$,    
B.~Abbott$^\textrm{\scriptsize 128}$,    
D.C.~Abbott$^\textrm{\scriptsize 102}$,    
O.~Abdinov$^\textrm{\scriptsize 13,*}$,    
A.~Abed~Abud$^\textrm{\scriptsize 70a,70b}$,    
K.~Abeling$^\textrm{\scriptsize 53}$,    
D.K.~Abhayasinghe$^\textrm{\scriptsize 93}$,    
S.H.~Abidi$^\textrm{\scriptsize 167}$,    
O.S.~AbouZeid$^\textrm{\scriptsize 40}$,    
N.L.~Abraham$^\textrm{\scriptsize 156}$
% ...
\bigskip
\\

% ...
$^{13}$Institute of Physics, Azerbaijan Academy of Sciences, Baku; Azerbaijan.\\
% ...
$^{40}$Niels Bohr Institute, University of Copenhagen, Copenhagen; Denmark.\\
% ...
$^{53}$II. Physikalisches Institut, Georg-August-Universit\"{a}t G\"ottingen, G\"ottingen; Germany.\\
% ...
$^{70}$$^{(a)}$INFN Sezione di Pavia;$^{(b)}$Dipartimento di Fisica, Universit\`a di Pavia, Pavia; Italy.\\
% ...
$^{93}$Department of Physics, Royal Holloway University of London, Egham; United Kingdom.\\
% ...
$^{101}$CPPM, Aix-Marseille Universit\'e, CNRS/IN2P3, Marseille; France.\\
$^{102}$Department of Physics, University of Massachusetts, Amherst MA; United States of America.\\
% ...
$^{128}$Homer L. Dodge Department of Physics and Astronomy, University of Oklahoma, Norman OK; United States of America.\\
% ...
$^{156}$Department of Physics and Astronomy, University of Sussex, Brighton; United Kingdom.\\
% ...
$^{167}$Department of Physics, University of Toronto, Toronto ON; Canada.\\
% ...
$^{a}$ Also at Borough of Manhattan Community College, City University of New York, New York NY; United States of America.\\
$^{b}$ Also at Centre for High Performance Computing, CSIR Campus, Rosebank, Cape Town; South Africa.\\
% ...
$^{*}$ Deceased

\end{flushleft}
